\documentclass[11pt,a4paper]{article}
\pdfoutput=1
\usepackage{jheppub}
\usepackage[utf8]{inputenc}
\usepackage{color,graphicx,hyperref,dsfont,slashed,multirow}
\hypersetup{bookmarks=true,unicode=true,pdftoolbar=true,pdfmenubar=true,
  pdffitwindow=false,pdfstartview={FitH},pdftitle={UMSSM},
  pdfauthor={J.~Araz,~G.~Corcella,~M.~Frank,~B.~Fuks}, pdfsubject={Subject},
  pdfcreator={Creator},pdfproducer={Producer}, pdfkeywords={Non-minimal 
  supersymmetry}{Beyond the Standard Model Physics},
  pdfnewwindow=true,colorlinks=true,
  linkcolor=blue,citecolor=magenta,filecolor=magenta,urlcolor=cyan}
\newcommand{\be}{\begin{equation}}
\newcommand{\ee}{\end{equation}}
\def\bsp#1\esp{\begin{split}#1\end{split}}

\newcommand\bpm{\begin{pmatrix}}
\newcommand\epm{\end{pmatrix}}
\def \lsim
{\raisebox{-3pt}{$\>\stackrel{<}{\scriptstyle\sim}\>$}}
\def \gsim
{\raisebox{-3pt}{$\>\stackrel{>}{\scriptstyle\sim}\>$}}

\def\sectionautorefname~#1\null{Sec.~#1\null}
\def\subsectionautorefname~#1\null{sub--Sec.~#1\null}
\def\figureautorefname~#1\null{Fig.~#1\null}
\def\tableautorefname~#1\null{Table~#1\null}
\def\equationautorefname~#1\null{Eq.~(#1)\null}

\date{\today}
\preprint{CUMQ/HEP 196}

\title{Loopholes in $Z^\prime$ searches at the LHC:\\
  exploring supersymmetric and leptophobic scenarios}

\author[a]{Jack Y. Araz}
\author[b]{\!, Gennaro Corcella}
\author[a]{\!, Mariana Frank}
\author[c,d,e]{\!and Benjamin Fuks}

\emailAdd{jack.araz@concordia.ca}
\emailAdd{gennaro.corcella@lnf.infn.it}
\emailAdd{mariana.frank@concordia.ca}
\emailAdd{fuks@lpthe.jussieu.fr}

\affiliation[a]{Department of Physics,
 Concordia University \\7141 Sherbrooke St. West, Montreal, QC,
 Canada H4B 1R6}
\affiliation[b]{INFN, Laboratori Nazionali di Frascati,\\
 Via E. Fermi 40,
 I-00044 Frascati (RM), Italy}
\affiliation[c]{Sorbonne Universit\'es, Universit\'e Pierre et Marie Curie\\
  (Paris 06), UMR 7589, LPTHE, F-75005 Paris, France}
\affiliation[d]{CNRS, UMR 7589, LPTHE,\\ F-75005, Paris, France}
\affiliation[e]{Institut Universitaire de France,\\ 103 boulevard Saint-Michel,
  75005 Paris, France}

\abstract{
  Searching for heavy vector bosons $Z^\prime$, predicted in models inspired by
  Grand Unification Theories, is among the challenging objectives of the LHC.
  The ATLAS and CMS collaborations have looked for $Z^\prime$ bosons assuming
  that they can decay only into Standard Model channels, and have set exclusion
  limits by investigating dilepton, dijet and, to a smaller extent, top-antitop
  final states. In this work we explore possible loopholes in these $Z^\prime$
  searches, by studying
  supersymmetric as well as leptophobic scenarios. We demonstrate the existence
  of realizations in which the $Z^\prime$ boson automatically evades the typical
  bounds derived from the analyses of the Drell--Yan invariant-mass spectrum.
  Dileptonic final states can in contrast only originate from supersymmetric
  $Z^\prime$ decays and are thus accompanied by additional effects. This feature
  is analyzed in the context of judiciously chosen benchmark configurations, for
  which visible signals could be expected in future LHC data with a
  $4\sigma-7\sigma$ significance. Our results should hence motivate an extension
  of the current $Z^\prime$ search program to account for supersymmetric and
  leptophobic models.}

\keywords{heavy neutral gauge boson, supersymmetry, UMSSM, LHC}

\begin{document}
\maketitle
\flushbottom


\section{Introduction}
\label{sec:intro}

Although the discovery of the Higgs boson at the LHC has allowed for the completion of the particle spectrum of the Standard Model (SM), the issue of
its extension still stands. Despite the experimental success in predicting
most data observed so far, the SM indeed exhibits several limitations and
shortcomings that motivate the study of beyond the Standard Model theories.
Among those, supersymmetry, and in particular its minimal incarnation known as
the Minimal Supersymmetric Standard Model (MSSM), is one of the most appealing
options. Unifying internal and external symmetries, supersymmetry provides
a natural solution to the long-standing hierarchy problem, allows for
gauge-coupling
unification at high energies and predicts a stable particle that could
address the problematics of Dark Matter. Despite these numerous motivations,
no compelling evidence for supersymmetry has been found and
the MSSM starts to be heavily
constrained. Moreover, the MSSM suffers from severe fine-tuning
issues related to the discovery of a SM-like Higgs boson, as
well as the
lack of any satisfactory explanation for the magnitude of the supersymmetric
bilinear Higgs mass parameter $\mu$
that must unnaturally be of the order of the
electroweak symmetry breaking scale.

As a consequence, arguments have been raised in favor of extending the MSSM
superfield content by at least one singlet chiral superfield. Its scalar
component can induce both supersymmetry
breaking and dynamical generation of the $\mu$ term by
getting a non-vanishing vacuum expectation value at the minimum of the
scalar potential~\cite{Kim:1983dt,Suematsu:1994qm,Cvetic:1996mf,Jain:1995cb,%
 Demir:1998dm}. Such singlet superfields also appear
under supersymmetric scenarios where the Standard Model gauge group is extended,
the scalar singlet yielding the breaking of the additional gauge
symmetry~\cite{Cvetic:1997ky,Demir:2005ti}. This setup is furthermore motivated
in a grand-unified scheme where a restricted set of high-dimensional
representations are used to encompass all MSSM supermultiplets, and where all
gauge couplings unify. In this context, the necessity of using
representations of the unified
gauge group automatically leads to the introduction of right-handed
neutrino superfields, which
consequently provides a solution for neutrino-mass generation, as well as
vector-like fermions.

Among all Grand Unification Theories (GUT), those based on gauge groups of
rank 6, named $E_6$, 
have been extensively discussed as interesting possibilities~\cite{%
 Langacker:2008yv,Demir:2005ti,Deppisch:2007xu}. In particular, the breaking
pattern of $E_6$ to the electroweak symmetry results in
the appearance of extra $U(1)^\prime$ symmetries. From a bottom-up perspective,
extending the MSSM with the introduction of an extra
$U(1)^\prime$ gauge group has numerous advantages, namely forbidding a
too rapid proton decay without introducing an {\it ad hoc} discrete
$R$-parity symmetry and making all field masses stable with respect to quantum
corrections. Moreover, it is always possible to choose the $U(1)^\prime$
field charges to ensure anomaly cancellation and gauge-coupling
unification. Besides, the $U(1)^\prime$ models do not suffer from the presence
of cosmological domain walls, unlike theories like the Next-to-Minimal
Supersymmetric Standard Model~\cite{Ellis:1986mq,Ellwanger:2009dp}.

While the spectrum of $U(1)^\prime$ supersymmetric (UMSSM)
models is altered from that of the MSSM, the most secure prediction emerging
from the extended gauge symmetry consists of the existence of a
novel neutral
$Z^\prime$ boson, like in non-supersymmetric $U(1)^\prime$ extensions of the SM. This
makes the $Z^\prime$ boson a prime target for the LHC physics program, as the proof of
its existence would constitute a promising indicator of a more general gauge
structure.
Any gauge group of rank greater than four (any group larger than
$SU(5)$) indeed leads to the appearance of at least one extra neutral gauge
boson. All current $Z^\prime$ analyses at the LHC are however guided by
non-supersymmetric considerations in which the $Z^\prime$ boson only decays into
SM particles~\cite{delAguila:2010mx,Salvioni:2009mt,Leike:1998wr,Cvetic:1995zs,%
 delAguila:1993ym}. Besides
$E_6$-inspired $Z^\prime$, the experimental collaborations
have also explored the so-called Sequential Standard
Model (SSM), the simplest extension of the Standard Model, wherein
$Z^\prime$ and possible $W^\prime$ bosons have the same couplings
to fermions as the $Z$ and $W$. This model is not theoretically
motivated, but it is often used as a benchmark for the analyses,
since the production cross section in the SSM
just depends on the extra boson masses.

Along these lines, the ATLAS and CMS collaborations
have searched for $Z^\prime$ bosons
by investigating dilepton and dijet final states.
In detail,  
by using high-mass dilepton data at 13 TeV, 
the ATLAS collaboration \cite{Aaboud:2017buh} set the mass exclusion limits
$M_{Z^\prime}>4.5$~TeV in the SSM and $M_{Z^\prime}>3.8$-4.1~TeV in
U(1)$'$ models, whereas CMS obtained $M_{Z^\prime}>4.0$~TeV (SSM) and   
$M_{Z'}>3.5$~TeV (GUT-inspired models) \cite{CMS:2016abv}.
For dijets, the limits are much milder and
read $M_{Z'}>2.1$-2.9~TeV (ATLAS) \cite{Aaboud:2017yvp}
and $M_{Z'}>2.7$~TeV (CMS) \cite{Sirunyan:2016iap}.

In a UMSSM framework, the inclusion of the
supersymmetric decay modes of the $Z'$ bosons may nonetheless change these
conclusions \cite{Gherghetta:1996yr,Baumgart:2006pa,
 Chang:2011be,Corcella:2012dw,
 Corcella:2014lha,Araz:2017qcs}.
Above all, the opening of new decay channels lowers the branching ratios
into SM final states and therefore the $Z^\prime$ mass exclusion limits.
In fact, Ref.~\cite{Corcella:2013cma} 
found an impact of about 200 GeV on the mass exclusion limits
by comparing the 8 TeV ATLAS and CMS data on high-mass dileptons with
UMSSM predictions for a benchmark point of the parameter space.
Furthermore, in the UMSSM, a leptophobic $Z^\prime$ can yield the
production of dilepton final states only 
through cascade decays into intermediate
electroweakinos, which contrasts with the leptophobic non-supersymmetric
case where this is simply not allowed~\cite{delAguila:1986ad}.
The bounds on the
$Z^\prime$-boson
mass and production cross section derived from the above-mentioned searches
should then be revisited when more general theoretical contexts like
the UMSSM or leptophobia are considered.

On different grounds, the hadronic environment at the LHC is so complex that new
physics searches
always rely on some simplifying assumptions in the form of the potential
signals. For instance, most supersymmetry searches have been designed from the
idea on how the MSSM could manifest itself in a typical LHC detector: they may
hence be not suitable for given non-minimal supersymmetric realizations. In the
UMSSM framework, which we focus on in this work, we consider
$Z^\prime$-boson signals that can potentially differ from the
non-supersymmetric case. We restrict our analysis to leptonic $Z^\prime$
decay modes that are easier to explore, even if the expected signals are plagued
by larger SM backgrounds. We additionally focus on UMSSM realizations in which
the $Z^\prime$ boson is leptophobic, but where it could give rise to leptonic
signatures via supersymmetric cascade decays into leptons and missing energy.
This therefore offers
an alternative opportunity to find both an extra gauge boson and supersymmetry
from the study of the decays of a resonantly-produced colorless particle. This
is one of the scenarios that we wish to investigate in this work, after imposing
the most up-to-date constraints on the model.
We hence aim at providing a clear roadmap for the discovery of
unconventional leptophobic $Z^\prime$ bosons,
such as those that could arise in UMSSM
scenarios and that escape detection when only considering standard LHC
searches for extra gauge bosons.

Our work is organized as follows. In Section~\ref{sec:model}, we briefly
introduce $U(1)^\prime$ supersymmetric models as when the gauge symmetry is
designed as emerging from the breaking of an extended $E_6$ symmetry at the
grand unification scale. We pay particular attention to the mass, mixing
patterns and interactions of the extra neutral gauge boson and show under what
conditions it could be made leptophobic. We finally set up the parameter-space
region to be scanned over and proceed to its exploration in
Section~\ref{sec:zprime}, focusing on two different way to impose boundary
conditions. In Section~\ref{sec:leptophobic}, we concentrate on
scenarios where the $Z^\prime$ boson does not directly decay into leptons and
study its phenomenology at colliders, highlighting a preferred selection
strategy that could lead to its discovery. We summarize our results and conclude
in Section~\ref{sec:conclusion}.

\section{$Z^\prime$ bosons in $U(1)^\prime$ supersymmetric models}
\label{sec:model}

\subsection{Theoretical framework}
\label{subsec:u1prime}

\begin{table}
 \renewcommand{\arraystretch}{1.5}
 \begin{center}
  \begin{tabular}{c||c|c|c|c|c|c}
     Model & $U(1)^\prime_\chi$ & $U(1)^\prime_\psi$ & $U(1)^\prime_\eta$ &
       $U(1)^\prime_S$ & $U(1)^\prime_I$ & $U(1)^\prime_N$\\
    \hline\hline
     $\theta_{E_6}$ & $-0.5\pi$ & $0$ & $-0.79\pi$  & $-0.37\pi$ & $0.71\pi$ &
       $-0.08\pi$
  \end{tabular}
  \caption{Mixing angle $\theta_{E_6}$ for the most popular
   $U(1)^\prime$ models. The value of $\theta_{E_6}$
   is imposed to lie in the $[-\pi,\pi]$ range.}
  \label{tab:theta6}\vspace{.3cm}

  \renewcommand{\arraystretch}{1.2}
  \begin{tabular}{l||c|c|c|c|c|c}
   & $2\sqrt{10}Q^{'}_\chi$ & $2 \sqrt{6}Q^{'}_\psi$ & $2\sqrt{15}Q^{'}_\eta$
    & $ 2\sqrt{15}Q^{'}_S $ & $ 2Q^{'}_I $ & $ 2\sqrt{10}Q^{'}_N $ \\
   \hline \hline
   $ Q,U,E$ & -1 & 1  & -2 & -1/2 & 0  & 1 \\
   $ L,D$   &  3 & 1  & 1  &   4  & -1 & 2 \\
   $ N $    & -5 & 1  & -5 &  -5  & 1  & 0 \\
   $ H_u$   & 2  & -2 & 4  &  1   & 0  & 2 \\
   $ H_d$   & -2 & -2 & 1  & -7/2 & 1  & -3\\
   $ S$       & 0  & 4  & -5 &  5/2 & -1 & 5 \\
  \end{tabular}
  \caption{$U(1)^\prime$ charges of the UMSSM quark ($Q$, $D$, $U$),
   lepton ($L$, $E$, $N$) and Higgs ($H_u$, $H_d$, $S$)
   supermultiplets for commonly studied anomaly-free $U(1)^\prime$
   groups that arise from the breaking of an $E_6$ symmetry.}
  \label{tab:u1charges}
 \end{center}
\end{table}

There are different ways to implement a $U(1)^\prime$ extension in the MSSM:
one of the most commonly used parameterizations is inspired by 
grand-unified models, based on a rank-6 group $E_6$, 
where the symmetry-breaking scheme proceeds via multiple steps,
\be\bsp
E_6 &\ \to SO(10) \otimes U(1)_\psi
\to SU(5) \otimes U(1)_\chi \otimes U(1)_\psi
\\ &\ \to SU(3)_C\otimes SU(2)_L \otimes U(1)_Y  \otimes U(1)^\prime \ .
\label{eq:e6brk}
\esp\ee

The $U(1)'$ symmetry that survives at the electroweak scale is taken as a linear
combination of $U(1)_\chi$ and $U(1)_\psi$,
\be
U(1)^\prime =\cos \theta_{E_6} U(1)_\psi - \sin\theta_{E_6} U(1)_\chi \ ,
\label{eq:sinthetaE6} \ee
where we have introduced the $E_6$ mixing angle $\theta_{E_6}$. The neutral
vector bosons associated with the $U(1)_\psi$ and $U(1)_\chi$ gauge groups are
called the $Z^\prime_\psi$ and $Z^\prime_\chi$ bosons, while a generic
$Z^\prime$ is given by the mixing of these $Z^\prime_\psi$ and $Z^\prime_\chi$
states, as in Eq.~\eqref{eq:sinthetaE6}.

Different $U(1)^\prime$ models can be classified according to the sole value of
the $\theta_{E_6}$ mixing angle, and the charges $Q^\prime$ of the
supermultiplets are fixed to ensure the theory to be anomaly-free. Six popular
setups are summarized in Table~\ref{tab:theta6}, with the corresponding
$Q^\prime$ charges listed in Table~\ref{tab:u1charges}. In the notations of this
last table, $Q$ and $L$ denote the left-handed weak doublets of quark and lepton
fields, $H_u$ and $H_d$ the two weak doublets of Higgs fields, $U$ and $D$ the
right-handed weak singlets of up-type and down-type quarks, $E$ and $N$ the
right-handed weak singlets of charged leptons and neutrinos, and $S$ a scalar
singlet. In the case of supersymmetric extensions of the Standard Model, such as
the MSSM, all fields in Table~\ref{tab:u1charges} must actually be understood as
superfields containing also the supersymmetric partners of the fermions and
Higgs bosons. 
In principle, the matter sector of $E_6$ should also feature
vector-like exotic (s)quarks $Q_D$ and $\bar Q_D$ which have the same
$U(1)^\prime$
charges as the $H_u$ and $H_d$ fields, respectively~\cite{Langacker:2008yv}.
In the following, we assume that these exotic states are 
too heavy to be relevant at LHC energies and neglect 
them in our phenomenological analysis\footnote{Due to the requirement of the
$SU(3)_c-SU(3)_c-U(1)^\prime$ anomaly cancellation, these exotic quarks have
weak isospin quantum numbers allowing for a superpotential interaction term
involving ordinary quarks and inducing rapid proton decay.
Their mass must thus be comparable to the GUT scale to prevent the proton from
decaying too quickly~\cite{Langacker:2008yv}.}.

The Higgs supermultiplet content
($H_u$, $H_d$ and $S$) is large enough to allow both for the breaking of 
$U(1)^\prime$ via the scalar singlet field $s$, and of the 
electroweak symmetry through the neutral components of the scalar Higgs doublets
$h_u$ and $h_d$. All electrically-neutral Higgs fields indeed get
non-vanishing vacuum expectation values at the minimum of the potential
and carry non-trivial $U(1)^\prime$ charges.

In the grand--unified framework, the
field content is organized into vector representations ($\mathbf{27}$) of the
$E_6$ group; the latter further branches as 
$\mathbf{27}=\mathbf{16}\oplus\mathbf{\overline{10}}\oplus
\mathbf{1}$ into the irreducible representations of the
$SO(10)$ subgroup that arises at the first step of the
$E_6$ breaking scheme of Eq.~\eqref{eq:e6brk}.
In the conventional field assignment, the representation $\mathbf{16}$
contains the left-handed quark and lepton supermultiplets ($Q$ and
$L$), as well as the right-handed quarks and leptons
($U$, $D$, $E$ and $N$), while the
Higgs fields ($H_u$ and $H_d$) and the exotic quarks $Q_D$ and $\bar Q_D$
are in the representation $\mathbf{10}$. An alternative framework
consists of having instead $H_u$ and $\bar Q_D$ lying in the $\mathbf{16}$ and
$L$ and $D$ in
the $\mathbf{10}$ representation. According to whether one chooses the
standard or unconventional assignment, the phenomenology of the $Z^\prime$
boson may be different. In the following, we shall adopt the standard $SO(10)$
representation choices, with the exotic quarks lying in the {\bf 10}
representation. Nevertheless, the unconventional scenario can be easily
recovered by redefining $\theta_{E_6}\to \theta_{E_6}+\arctan{\sqrt{15}}$
in Eq.~(\ref{eq:sinthetaE6})~\cite{Nardi:1994ty}.

In principle, the Higgs fields in the {\bf 27} representation
of $E_6$ should occur
in three generations.
However, as discussed in Refs.~\cite{Hewett:1988xc,Ellis:1986ip,Georgi:1978ri},
it is always possible to perform a unitary transformation to a basis
where only one generation of Higgs bosons gets a
non-vanishing vacuum expectation value. The scalars with zero
vacuum expectation values were called `unHiggs' in
Refs.~\cite{Hewett:1988xc,Ellis:1986ip}.
Through our analysis, 
we shall neglect the two generations of such states and focus on 
the `true'
Higgs bosons, which exhibit a non-zero vacuum expectation values and are
denoted by $H_u$ and $H_d$.

The $\mathbf{16}$ representation of $SO(10)$
is then decomposed in terms of those of $SU(5)$
as $\mathbf{16}=\mathbf{10}\oplus\mathbf{\overline{5}}\oplus
\mathbf{1}$.
The $\mathbf{10}$ representation of $SU(5)$ is suitable to include right-handed
up-type quark and charged-lepton supermultiplets, together with the weak
doublets of left-handed quarks, whereas the $\mathbf{\overline{5}}$
representation contains right-handed down quarks and left-handed lepton
supermultiplets; the $\mathbf{1}$ representation includes right-handed
(s)neutrinos~\cite{Slansky:1981yr}. The UMSSM
superpotential is thus given, all flavor indices being omitted for clarity, by:
\be
W_{\rm UMSSM} = U\,{\mathbf Y}_u\,Q\,H_u\, - D\,{\mathbf Y}_d \,Q\,H_d\,-
  E\,{\mathbf Y}_e \,L\,H_d\,+ N\,{\mathbf Y}_\nu\,L\,H_u\,+
  \lambda\,H_u\,H_d\,S \ .
\ee
The Yukawa
interactions are encoded in a set of four $3\times 3$ matrices in flavor space,
${\mathbf Y}_u$, ${\mathbf Y}_d$, ${\mathbf Y}_l$ and ${\mathbf Y}_\nu$, and the
strength of the supersymmetric Higgs self-interactions is described by the
$\lambda$ parameter. After the breaking of the $U(1)^\prime$ symmetry, this
$\lambda$-term induces the dynamical generation of an effective $\mu$-term
(denoted $\mu_{\rm eff}$ in the following) that allows for the
resolution of the so-called MSSM $\mu$-problem\footnote{$\mu_{\rm eff}$ is
related to $\lambda$ and to the vacuum expectation
value of the scalar singlet $s$ via $\mu_{\rm eff}=\lambda \langle s\rangle$.}.
As supersymmetry has to be
softly broken, we introduce in the Lagrangian explicit mass terms for all
gaugino and scalar fields,
\be\bsp
& {\cal L}_{\rm soft}^{({\rm masses})} = \frac12 \Big(
M_1  \lambda_{\tilde B} \!\cdot\! \lambda_{\tilde B} +
M_2  \lambda_{\tilde W} \!\cdot\! \lambda_{\tilde W} +
M_3  \lambda_{\tilde g} \!\cdot\! \lambda_{\tilde g} +
M_4 \lambda_{\tilde B'}\!\cdot\! \lambda_{\tilde B'}+
{\rm h.c.}\Big) 
- m_{H_d}^2h_d^\dag h_d \\ &\quad - m_{H_u}^2 h_u^\dag h_u -
\frac12 m_{s}^2 s^2 - m^2_{\tilde q}{\tilde q}^\dag\tilde q -
m^2_{\tilde d}{\tilde d}^\dag{\tilde d} -
m^2_{\tilde u}{\tilde u}^\dag\tilde u -
m^2_{\tilde l}{\tilde l}^\dag \tilde l -
m^2_{\tilde e}{\tilde e}^\dag{\tilde e}-
m^2_{\tilde \nu}{\tilde \nu}^\dag{\tilde \nu} \ ,
\esp\ee
where the $U(1)_Y$, $U(1)'$, $SU(2)_L$ and $SU(3)_c$ gaugino Weyl fermions are
denoted by $\lambda_{\tilde B}$, $\lambda_{\tilde B'}$,
$\lambda_{\tilde W}$ and $\lambda_{\tilde g}$, respectively, and where $h_d$,
$h_u$, $s$, $\tilde q$, $\tilde d^\dag$, $\tilde u^\dag $, $\tilde l$,
$\tilde e^\dag$ and $\tilde\nu^\dag$ are the scalar components of the $H_d$,
$H_u$, $S$, $Q$, $D$, $U$,
$L$, $E$ and $N$ superfields. The set of $M_i$ and $m_i$ parameters moreover
denote the soft gaugino and scalar mass parameters, respectively.

Additional soft terms, related to trilinear
scalar interactions, are also present and can be derived from the structure of
the superpotential,
\be
  {\cal L}_{\rm soft}^{({\rm tril.})} = - A_\lambda\, s\, h_u\, h_d\, +
  {\tilde d}^\dag\,{\mathbf A_d}\, {\tilde q}\, h_d\, +
  {\tilde e}^\dag\,{\mathbf A_e}\, {\tilde l}\, h_d\, -
  {\tilde u}^\dag\,{\mathbf A_u}\, {\tilde q}\, h_u\, -
  {\tilde \nu}^\dag\,{\mathbf A_\nu}\, {\tilde l}\, h_u\, +{\rm h.c.}\ ,
\ee
where the ${\mathbf A}_e$, ${\mathbf A}_\nu$, ${\mathbf A}_d$ and
${\mathbf A}_u$ $3 \times 3$ matrices stand for the strengths of the soft Higgs-boson
interactions with charged sleptons, sneutrinos, down-type
squarks and up-type squarks, respectively. The $A_\lambda$ parameter is finally
related to the trilinear soft multi Higgs-boson coupling.

In order to calculate the sfermion masses, one would need
to set up an explicit framework for supersymmetry breaking,
such as a gauge-, gravity- or anomaly-mediated mechanisms, which
goes beyond the goals of the present paper. We only recall
that supersymmetry can be spontaneously broken if the so-called
$D$-term and/or $F$-term in the scalar potential have non-zero
vacuum expectation values. The $F$-terms are proportional
to the SM particle masses, and are therefore important only
for stop quarks, whereas $D$-terms 
are relevant for both light
and heavy sfermions and contain contributions due to electroweak
symmetry breaking and, in case of extension of the MSSM, to the Higgs bosons
which break the extended symmetry~\cite{Gherghetta:1996yr,Corcella:2012dw,%
Corcella:2014lha}.
Hereafter, we account for $F$- and $D$-term corrections to the
sfermion masses, but do not present
their explicit expressions, for the sake of brevity.

After the spontaneous
breaking of the symmetry group down to electromagnetism,
the $W$, $Z$ and $Z^\prime$ bosons get massive and the photon stays massless.
In general, for a $U(1)^\prime$ extension of the SM, there is mixing
between the $Z$ and $Z^\prime$ eigenstates, parameterized by
a mixing angle $\alpha_{ZZ^\prime}$. However,
electroweak precision data strongly constrain  $\alpha_{ZZ^\prime}$
to be very small \cite{Erler:2002pr}.
At tree level, the squared masses of the $Z$ and $Z'$ bosons are given by:
\begin{eqnarray}\label{mzz}
M_Z^2 &=& \frac{g_1^2+g_2^2}{2}
\bigg( \langle h^0_u \rangle^2 + \langle h^0_d \rangle^2\bigg)\nonumber\\
M_{Z^\prime}^2 &=& g^{\prime 2}
\bigg({Q^\prime_S}^2 \langle s \rangle^2 + {Q^\prime_{H_u}}^2 \langle h^0_u
\rangle^2 +
{Q^\prime_{H_d}}^2\langle h^0_d \rangle^2\bigg) \ ,
\end{eqnarray}
where $h_d^0$ and $h_u^0$ stand for the neutral components of the down-type and
up-type Higgs fields $h_d$ and $h_u$ and $g_1$, $g_2$ and $g^\prime$ are the
coupling constants of the $U(1)_Y$, $SU(2)_L$ and $U(1)^\prime$ gauge
groups, respectively. As discussed, {\it e.g.}, in Ref.~\cite{Langacker:2008yv},
whenever the singlet $s$ has a large vacuum expectation value (which
contributes only to the $Z^\prime$ mass), as will be the case hereafter,
$M_Z^2\ll M_{Z'}^2$.

In the Higgs sector, as discussed above, one should deal
with three generations of Higgs fields, although, in our chosen basis,
only one generation (the so-called `true' Higgs bosons) exhibits non-zero
vacuum expectation values.
Mass mixing matrices and mass eigenstates of the two generations of
Higgs bosons with zero vacuum expectation values are thoroughly
debated in \cite{Ellis:1986ip}. In principle, because of the presence of
these other states, one should impose 
further constraints on our scenario
coming, e.g., from the current measurements
of the (SM-like) neutral-Higgs production cross section and branching
ratios, as well as from the exclusion limits on charged-Higgs bosons.
In our work, however, such extra Higgs states and related
constraints will be neglected.

In fact, after electroweak symmetry breaking,
for each generation of Higgs fields,
one is left with two charged and four neutral
scalar bosons, namely  one pseudoscalar and 
three
neutral scalars, including a novel
singlet-like scalar Higgs, inherited
by the $U(1)^\prime$ symmetry.
In the following, we shall account for only one generation of Higgs bosons
and denote by  $H^\pm$ the charged bosons, $h$ and $H$ the
MSSM-like neutral scalars, with $h$ roughly corresponding to
the Standard Model Higgs,
$A$ the pseudoscalar and  $H^\prime$ the
extra scalar associated with the $U(1)^\prime$ gauge group.


As discussed, e.g., in Ref.~\cite{Hewett:1988xc}, for
$\langle s\rangle$ much larger than $ \langle h^0_u \rangle$
and $\langle h^0_d \rangle$, diagonalizing the
neutral Higgs mass matrix is straightforward and the singlet-like $H^\prime$
has mass
$M_{H^\prime}^2\simeq g^{\prime 2}{Q^\prime_S}^2\langle s\rangle^2$, hence
it is roughly degenerate with the $Z^\prime$, according to Eq.~(\ref{mzz}).
The other neutral Higgs $H$ has instead approximately the same mass as the
pseudoscalar $A$ and as the charged $H^\pm$:
as a result, the heaviest scalar Higgs of the
spectrum could be either $H$ or $H^\prime$, depending on whether
the $Z^\prime$ is lighter or heavier than $A$.

In the gaugino sector, with respect to the MSSM, one has two extra neutralinos,
related to the supersymmetric partners of $Z^\prime$ and $H^\prime$ bosons,
which yields a total of six $\tilde\chi^0_1, \dots, \tilde\chi^0_6$ neutralino
states. As discussed in Ref.~\cite{Corcella:2014lha},
the new $\tilde\chi^0_5$ and $\tilde\chi^0_6$ eigenstates are often too heavy to
contribute to the $Z^\prime$ phenomenology at the LHC.
As the new $Z^\prime$ is electrically neutral, the chargino
sector stays instead unchanged with respect to the MSSM.

On top of mass mixings, both $U(1)_Y$ and $U(1)^\prime$ bosons are allowed to
mix kinetically~\cite{Babu:1997st}.
The corresponding Lagrangian reads, in terms of
the gauge boson component fields,
\be
{\cal L}_{\rm kin} =
- \frac14 \hat B^{\mu\nu} \hat B_{\mu\nu}
- \frac14 \hat Z^{\prime\mu\nu} \hat Z^\prime_{\mu\nu}
- \frac{\sin\chi}{2} \hat B^{\mu\nu} \hat Z^\prime _{\mu\nu} \ ,
\label{eq:kineticmix} \ee
where $\hat B_{\mu\nu}$ and $\hat Z^\prime_{\mu\nu}$ are the $U(1)_Y$ and
$U(1)'$ boson field strength tensors, respectively, and $\chi$ is
the kinetic mixing angle.
In order to understand the physical implications of the kinetic
mixing, it is necessary to diagonalize the field strengths, 
which is achieved via a $GL(2,\mathbb{R})$ rotation,
\be\label{bbzz}
\bpm \hat B_\mu\\ \hat Z^\prime_\mu \epm = 
\bpm 1 & -\tan \chi\\ 0 & \frac{1}{\cos \chi} \epm
\bpm B_\mu \\Z^\prime_\mu\epm \ ,
\ee
where $\hat B_\mu$ and $Z^\prime_\mu$ are the original
$U(1)$ and $U(1)^\prime$ gauge fields, with non-diagonal
kinetic terms, while 
$B_\mu$ and $Z^\prime_\mu$ have now canonical diagonal
kinetic terms.
As discussed in Refs.~\cite{Babu:1997st,Langacker:2008yv}, 
for  $M^2_{Z} \ll M^2_{Z^\prime}$
and small values of $\chi$, the impact of the kinetic mixing on
the gauge boson masses is negligible. It nonetheless
can have a significant effect on the coupling of the $Z^\prime$ boson with
fermions. In fact, the interaction Lagrangian of the fields $\hat B_\mu$ and
$\hat Z^\prime_\mu$ with a generic fermion $\psi_i$, with charges $Y_i$ and
$Q^\prime_i$ under the $U(1)$ and $U(1)^\prime$ groups, is given by
\be
{\cal L}_{\rm int}=-\bar\psi_i\gamma^\mu(g_1Y_i\hat B_\mu+g^\prime Q^\prime_i
\hat Z^\prime_\mu)\psi_i \ ,
\ee
which  can then be rewritten in terms of $B_\mu$ and $Z^\prime_\mu$ as
\be
{\cal L}_{\rm int}=-\bar\psi_i\gamma^\mu(g_1Y_iB_\mu+g^\prime\bar
Q_i Z^\prime_\mu)\psi,
\ee
where
\be
\bar Q_i=Q_i^\prime\sec\chi-\frac{g_1}{g^\prime}Y_i\tan\chi.
\label{eq:Qbardef}
\ee
Leptophobic scenarios can hence be obtained
requiring
$\bar Q_L=\bar Q_E =0$~\cite{Babu:1996vt,Suematsu:1998wm,Chiang:2014yva}.
Since $Y_L=-1/2$ and $Y_E=1$,  Eq.~\eqref{eq:Qbardef} dictates that
leptophobia can be achieved only if
$Q^\prime_E = -2Q^\prime_L$: this
relation between the doublet and singlet leptonic charges is typical for the
$U(1)^\prime_\eta$ configuration, as shown in Table~\ref{tab:u1charges}.
Furthermore, if one assumes, as will be done in the following,
the typical GUT-inspired relation between the $U(1)$ and $U(1)^\prime$
couplings $g_1 (M_{Z^\prime})/g^\prime (M_{Z^\prime})=\sqrt{3/5}$, then leptophobia requires the 
additional condition
$\sin\chi\approx-0.3$.
As a result, we expect leptophobic $Z^\prime$
models to naturally arise for $E_6$ mixing angles in the neighbourhood of
\be
\theta_{E_6} \simeq \theta_\eta \pm n\pi,\quad   n=0,1,2,3, \ldots\ , 
\label{eq:e6mixlept}\ee
with the $Z^\prime$-boson leptonic couplings being either exactly zero or very suppressed.
In the following, we shall account for the kinetic mixing of
$U(1)_Y$ and $U(1)^\prime$ gauge groups, with the $U(1)^\prime$ charges of all our
matter fields given by Eq.~(\ref{eq:Qbardef}).

\subsection{Parameter-space scan and constraints}
\label{subsec:scan}

UMSSM theories rely on numerous free parameters so that simplifying
assumptions are in order for a practical parameter-space exploration.
Hereafter, we impose
minimal flavor violation,
so that all the flavor-violating parameters of the soft
supersymmetry-breaking Lagrangian are considered as vanishing, and
enforce unification boundary conditions on the remaining soft parameters.

\begin{table}
 \renewcommand{\arraystretch}{1.2}
 \begin{center}
  \begin{tabular}{c|c||c|c}
   Parameter      & Scanned range& Parameter      & Scanned range\\
   \hline
   $M_0$          & $[0, 3]$~TeV & $\mu_{\rm eff}$          & $[-2, 2]$~TeV\\
   $M_{1/2}$      & $[0, 5]$~TeV & $A_\lambda$    & $[-7, 7]$~TeV\\
   $A_0$          & $[-3, 3]$~TeV& $M_{Z'}$       & $[1.98, 5.2]$~TeV\\
   $\tan\beta$    & $[0, 60]$    & $\theta_{E_6}$ & $[-\pi, \pi]$ 
  \end{tabular}\vspace*{.5cm}
  \begin{tabular}{c|c||c|c}
   Parameter      & Scanned range& Parameter      & Scanned range\\
   \hline
   $m^2_{\tilde q,\tilde u,\tilde d}$  & $[0, 16]\ ~$TeV$^2$ &
   $M_{1,2,3,4}$ & $[0, 3]$~TeV \\
   $m^2_{\tilde e,\tilde l}$    & $[0, 1]~$TeV$^2$   & $m^2_{\tilde\nu}$
   & $[-6.8,9]~$TeV$^2$\
  \end{tabular}
  \caption{Ranges over which we allow the parameters in
   Eqs~\eqref{eq:prms1} and \eqref{eq:prms2} to vary. As discussed in the
   text, for coupling unification at GUT scale, only the quantities in
   the top panel are varied. }
  \label{tab:scan_lim}
 \end{center}
\end{table}

In the first class of scenarios which we investigate,
unification is assumed to occur at a very high
scale $M_{\rm GUT}\approx {\cal O}(10^{16})~{\rm GeV}$
and all parameters are then run down to $M_{Z^\prime}$ 
according to renormalization group evolution. More precisely, all gauge
couplings are assumed to unify at a given high scale and the $U(1)^\prime$
coupling is enforced to satisfy
\be
g^\prime(M_{\rm GUT})=\sqrt{\frac{5}{3}}\ g_1(M_{\rm GUT}).
\ee
Furthermore, all scalar masses are set to a common value
$M_0$, whilst all gaugino
masses are taken equal to another universal mass $M_{1/2}$. All
trilinear soft couplings are assumed to be proportional to the
respective Yukawa coupling matrices with a universal proportionality factor
$A_0$, so that
\be
{\mathbf A}_i = {\mathbf Y}_i A_0 \qquad \text{for}~i=e, \nu, d,
u \ .
\label{eq:trilU}\ee
In the Higgs sector, we fix the values of the effective
$\mu_{\rm eff}$ parameter, the ratio of the vacuum expectation
values of the neutral components of the two Higgs doublets
$\tan\beta=v_u/v_d$, the
trilinear soft coupling  $A_\lambda$, as well as the $Z^\prime$ mass
$M_{Z^\prime}$.
Finally, the diagonal entries of the neutrino Yukawa coupling matrices are set
to a very small value, 
${\cal O} (10^{-11})$, in such a way as to ignore the sneutrino soft
trilinear interactions. The ensemble of free parameters considered in our
exploration of the UMSSM parameter space is thus given by
\be
\bigg\{
  M_0,\ M_{1/2},\ A_0,\ \tan\beta,\ \mu_{\rm eff},\ A_\lambda,\ M_{Z'},\
   \theta_{E_6}
\bigg\} \ ,
\label{eq:prms1}\ee
where we have additionally included the $E_6$ mixing angle $\theta_{E_6}$.
We vary those parameters over the ranges given in the top
panel of Table~\ref{tab:scan_lim}.

In the second class of scenarios considered in this work, unification is imposed
at the $Z^\prime$ mass scale. In this case, we just enforce
the unification of the trilinear couplings as in Eq.~\eqref{eq:trilU} and set
\be
g'(M_{Z^\prime}) = \sqrt{\frac{5}{3}}\ g_1(M_{Z^\prime}) \ ,
\ee
all scalar and gaugino masses being kept free. The entire set of free
parameters is thus here given by
\be\label{eq:prms2}
\bigg\{
  m^2_{\tilde q},\ m^2_{\tilde u},\ m^2_{\tilde d},\ m^2_{\tilde l},\
  m^2_{\tilde e},\ m^2_{\tilde \nu},\ M_1,\ M_2,\ M_3,\ M_4,\
  A_0,\ \tan\beta,\ \mu_{\rm eff},\ A_\lambda,\ M_{Z'},\ \theta_{E_6}
\bigg\} \ ,
\ee
with the ranges over which those parameters vary presented 
in Table~\ref{tab:scan_lim}.

\begin{table}
 \renewcommand{\arraystretch}{1.2}
 \begin{center}
  \begin{tabular}{l|c|c||l|c|c}
   Observable & Constraints & Ref. & Observable & Constraints & Ref.\\
   \hline\hline
   $M_h$ & $ 125.09 \pm 3 $ GeV (theo) & \cite{Chatrchyan:2012xdj} &
   $\chi^2(\hat{\mu})$ & $\leq 70 $ & - \\
   $|\alpha_{ZZ^\prime}| $& $ {\cal O}(10^{-3}) $ & \cite{Erler:2009jh} &
   $M_{\tilde{g}} $& $ > 1.75 $ TeV & \cite{Khachatryan:2016xvy}\\
   $M_{\tilde\chi^0_2}$ & $ > 62.4$ GeV & \cite{Agashe:2014kda} & 
   $M_{\tilde\chi^0_3} $ & $ > 99.9 $ GeV & \cite{Agashe:2014kda}\\
   $M_{\tilde\chi^0_4} $      & $ > 116 $ GeV & \cite{Agashe:2014kda} &
   $M_{\tilde\chi^\pm_i} $    & $ > 103.5 $ GeV & \cite{Agashe:2014kda}\\
   $M_{\tilde{\tau}} $ & $ > 81 $ GeV & \cite{Agashe:2014kda} &
   $M_{\tilde{e}} $ & $ > 107 $ GeV & \cite{Agashe:2014kda} \\
   $M_{\tilde{\mu}} $& $ > 94 $ GeV & \cite{Agashe:2014kda} &
   $M_{\tilde{t}} $& $ > 900 $ GeV & \cite{Aaboud:2016lwz}\\
   BR$(B^0_s \to \mu^+\mu^-) $ & $[1.1\times10^{-9},6.4\times10^{-9}]$ &
   \cite{Aaij:2012nna} & $\displaystyle  \frac{{\rm BR}(B \to \tau\nu_\tau)}
   {{\rm BR}_{SM}(B \to \tau\nu_\tau)} $ &$  [0.15,2.41] $ &
   \cite{Asner:2010qj}\\
   BR$(B^0 \to X_s \gamma) $ &$  [2.99,3.87]\times10^{-4} $ &
   \cite{Amhis:2016xyh}\\
  \end{tabular}
  \caption{\label{tab:constraints}Experimental constraints imposed within
   our scanning procedure in order to determine the parameter-space regions of
   interest.}
 \end{center}
\end{table}

In our scanning procedure, we analyze all possible anomaly-free UMSSM models
derived from the breaking of an $E_6$ gauge symmetry. We generate the particle
spectrum by making use of the \textsc{Sarah} code, 
version 4.6.0~\cite{Staub:2013tta}, and
its interface to \textsc{SPheno} 3.3.8~\cite{Porod:2011nf}. In order to
test the phenomenological viability of the model, we compute various
properties of the Higgs sector, such as the mass of the lightest Higgs state and
the corresponding collider signal strengths by means
of the \textsc{HiggsBounds} (version 4.3.1) and \textsc{HiggsSignals} (version
1.4.0) packages~\cite{Bechtle:2008jh,Bechtle:2013xfa}. The scan
itself and the numerical analysis performed in this work have been achieved by
interfacing all programs using also the \textsc{pySLHA} package, version
3.1.1~\cite{Buckley:2013jua}.

The parameter space is probed by using the Metropolis--Hasting sampling method,
requiring consistency with the experimental bounds on masses and
decay rates shown in
Table~\ref{tab:constraints}. In particular,
we require the mass of the Standard Model Higgs
boson to agree with the measurements up to an uncertainty of 3~GeV, and the
$\chi^2$ fit of the available Higgs signal strengths is bounded to be smaller
than the conservative value of 70. Other constraints,
connected to the bounds on the masses of supersymmetric particles
and on several flavor observables, 
are evaluated relying on
the {\sc SPheno} code. This includes in particular tests of the
strict limits stemming from $B$-meson
decays~\cite{Aaij:2012nna,Asner:2010qj,Aaboud:2016lwz}.
As for the supersymmetric sector,
we enforce the LEP limits on slepton, chargino, and neutralino masses
quoted in Ref.~\cite{Agashe:2014kda}, while for gluinos and stops
we implement the bounds set by 
CMS \cite{Khachatryan:2016xvy} and ATLAS
\cite{Aaboud:2016lwz}, respectively.


\section{Supersymmetric $Z^\prime$ Phenomenology}
\label{sec:zprime}

In this section, we analyze the phenomenology of the two classes of UMSSM
scenarios introduced in Section~\ref{subsec:scan}. In the subsequent
Section~\ref{sec:leptophobic}, specific configurations where
the $Z^\prime$ boson is leptophobic by virtue of
the kinetic mixing of $U(1)_Y$ and $U(1)^\prime$ are in contrast investigated.

In order to apply the LHC constraints on the properties of $Z^\prime$ bosons, we
calculate the $Z^\prime$ production cross section
at next-to-leading order
(NLO) accuracy in QCD~\cite{Fuks:2007gk,Fuks:2017vtl}. This relies on the joint
use of
\textsc{FeynRules} version 2.3.27~\cite{Alloul:2013bka} and the included \textsc{NLOCT}
package~\cite{Degrande:2014vpa}, as well as
{\sc FeynArts}~\cite{Hahn:2000kx}, for the automatic generation of a UFO
library~\cite{Degrande:2011ua} containing both tree-level and counterterm
vertices necessary at NLO. This UFO model is then used by
\textsc{MadGraph5\_aMC@NLO} (version 2.5.5)~\cite{Alwall:2014hca} for the
numerical evaluation of the hard-scattering matrix elements, which are
convoluted with the NLO set of NNPDF 2.3 parton distribution
functions (PDF)~\cite{Ball:2012cx}. Using the decay table provided by the
\textsc{SPheno} package and assuming the narrow-width approximation, we compare
our predictions with the ATLAS limits on $Z'$
bosons in the dilepton mode~\cite{Aaboud:2017buh} in order
to estimate the impact of supersymmetric decay channels. 

\subsection{Scenarios With High-Scale Boundary Conditions}
\label{subsec:td}
In this subsection, we focus on our first class of UMSSM scenarios where 
the proportionality between $g^\prime$ and $g_1$ is imposed at the GUT scale
and where all free
parameters in Eq.~\eqref{eq:prms1} are fixed at $M_{\rm GUT}$
and then evolved down to the $Z^\prime$ scale by means of renormalization
group equations.

\begin{table}
 \renewcommand{\arraystretch}{1.4}
 \begin{center}
  \begin{tabular}{c|cccc}
   Parameter & $U(1)^\prime_\psi$ & $U(1)^\prime_\eta$ &
   $U(1)^\prime_I$ & $U(1)^\prime_N$ \\\hline\hline
   $g^\prime_{\rm min}$   & 0.634 & 0.585 & 0.559 & 0.624 \\
   $\Delta g^\prime$ [\%] & 0.9  & 7.8  & 6.8  & 1.4 \\
   $[{\rm BR}(Z^\prime\to ll)]^{\rm min}_{\rm UMSSM}$
   [\%]& 5.5 & 3.6 & 9.3 & 7.8 \\
   $[{\rm BR}(Z^\prime\to ll)]^{\rm min}_{\rm USM}$
   [\%]& 8.4 & 4.8 & 11.1 & 11.1 \\
  \end{tabular}
  \caption{\label{tab:sce1assump} $g^\prime$ values and dilepton
   branching ratios for commonly studied $U(1)'$ models
   with UMSSM parameters satisfying the
   constraints detailed in subsection~\ref{subsec:scan}.
   Quoted are $g^\prime_{\rm min}$, 
   the minimum value of $g^\prime(M_{Z^\prime})$, along with 
   the corresponding spread
   $\Delta g^\prime$ and the smallest possible branching ratio into leptons
   with (UMSSM) and without (USM) supersymmetric contributions to the $Z^\prime$
   decays.}
 \end{center}
\end{table}

\begin{figure}
 \centering
 \includegraphics[width=0.7\columnwidth]{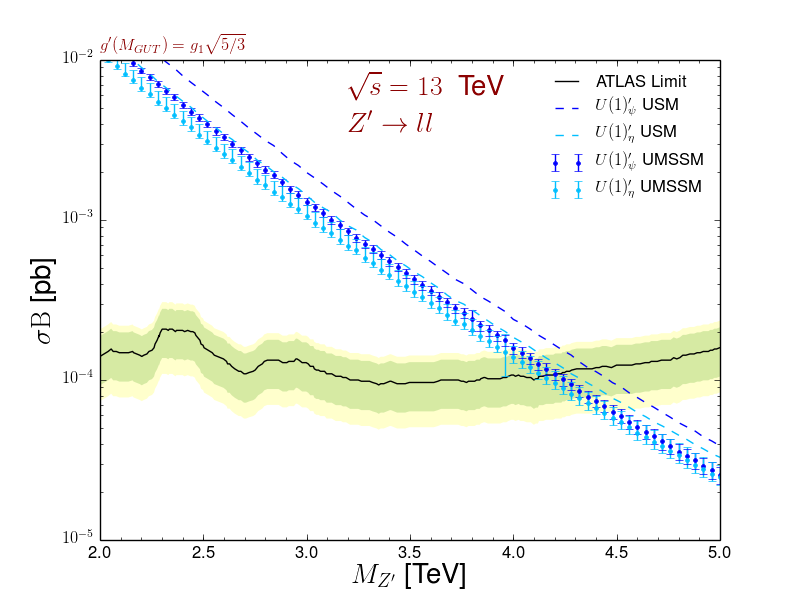}\\
 \includegraphics[width=0.7\columnwidth]{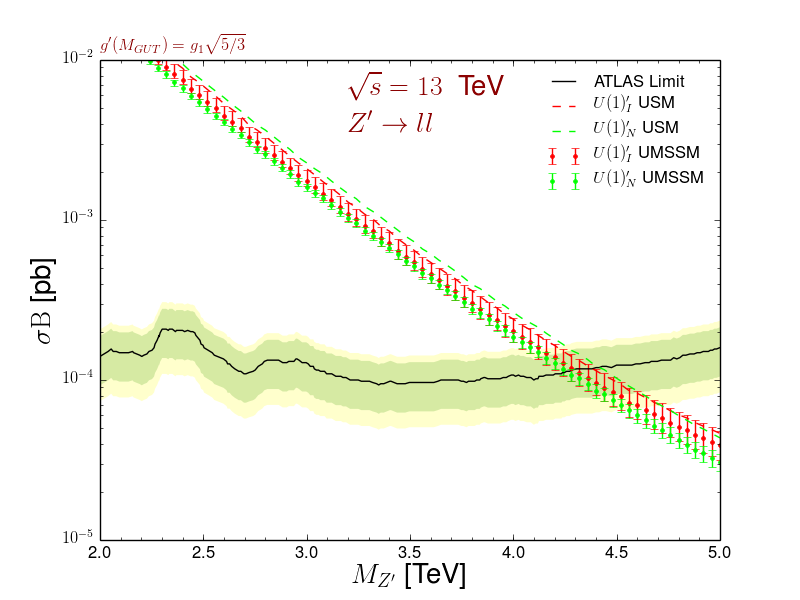}
 \caption{Comparison of our predictions for the
  $\sigma(pp\to Z^\prime)\times{\rm BR}(Z^\prime\to ll)$ product, in the
  scenario where the model boundary conditions are set at $M_{\rm GUT}$, 
  with the ATLAS dilepton yield~\cite{Aaboud:2017buh} at the $1\sigma$ (green)
  and $2\sigma$ (yellow) confidence levels.
  In the upper panel, we present the results for the $U(1)^\prime_\psi$ and
  $U(1)^\prime_\eta$ models, and in the lower panel we focus on the
  $U(1)^\prime_I$ and $U^\prime_N$ models.
  The dots with error bands
  correspond to the UMSSM case, while the dashed lines do not include
  supersymmetry (USM). NLO corrections to $\sigma(pp\to Z^\prime)$ are
  accounted for in both cases and the spread in the UMSSM results
  includes the effects of the parameter scan as well as the theoretical
  error originating from scale and PDF variations.}
 \label{fig:sigBbench1}
\end{figure}

We have found that some parameter regions satisfying the constraints
in Table~\ref{tab:constraints}
exist for a wide set of values of the
$E_6$ mixing angle $\theta_{E_6}$.
The LHC collaborations typically use the rate
$\sigma B \equiv \sigma(pp\to Z^\prime)\times {\rm BR}(Z^\prime\to l^+ l^-)$ to
obtain the exclusion limits on the $Z^\prime$ mass. For the sake of exploring
possible loopholes in the $Z^\prime$ searches, we are therefore especially
interested in scenarios which minimize the $\sigma{\rm B}$ product, namely
featuring small values of the $g^\prime$ coupling and of the
${\rm BR}(Z^\prime\to l^+ l^-)$ branching ratio.
In fact, when running the renormalization group equations,
scanning the parameters in the ranges presented in
Table~\ref{tab:scan_lim}, imposing the constraints of
Table~\ref{tab:constraints}
and accounting for proper threshold matching conditions,
$g^\prime(M_{Z'})$ ends up
with lying in a range $[g^\prime_{\rm min},g^\prime_{\rm max}]$.

In Table \ref{tab:sce1assump} we quote, for a few
$U(1)^\prime$ models, the minimum value of $g^\prime$ at the
$M_{Z^\prime}$ scale and
the spread $\Delta g^\prime$,
defined as
\begin{equation}
  \Delta g^\prime  =1-\frac{g^\prime_{\rm min}(M_{Z^\prime})}
         {g^\prime_{\rm max}(M_{Z^\prime})}
\end{equation}
and expressed as a percentage.
The
minimum branching fraction of  $Z^\prime$ decays into dilepton final states, including supersymmetric channels 
(UMSSM) and without supersymmetry (USM) is also
quoted.

In the table, we have discarded the models $U(1)^\prime_\chi$ and
$U(1)^\prime_S$. As discussed, {\it e.g.}, in Refs.~\cite{Corcella:2012dw,%
Araz:2017qcs}, $U(1)^\prime_\chi$ models are ill-defined in supersymmetry as it
they typically
lead to unphysical sfermion masses after adding to the soft masses
the $D$-term contributions.
As to $U(1)^\prime_S$, it may be theoretically acceptable, but
we were not able to find scenarios
capable of satisfying the constraints of Table~\ref{tab:constraints}.
From Table~\ref{tab:sce1assump}, we learn that the
deviations of $g^\prime$ 
from  $g^\prime_{\rm min} $ are rather small, with $\Delta g^\prime$ being
of at most about 8\%,
but the impact of the inclusion of supersymmetric decays on the
dilepton branching fraction
is remarkable for most models.
In the $U(1)^\prime_\psi$ and $U(1)^\prime_\eta$ scenarios, for example,
${\rm BR}(Z^\prime\to ll)$ decreases by about 35\% and
25\%, respectively, once decays into sfermions and gauginos are accounted
for. Nevertheless, all models still exhibit substantial dilepton $Z'$ decay
rates, varying between 3\% and 10\%.

\begin{figure}
 \centering
 \includegraphics[width=0.7\columnwidth]{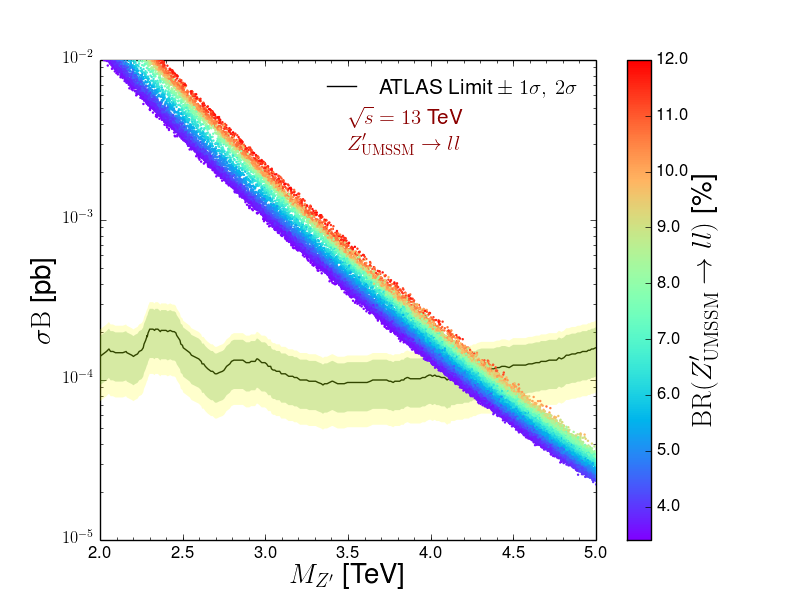}
 \includegraphics[width=0.7\columnwidth]{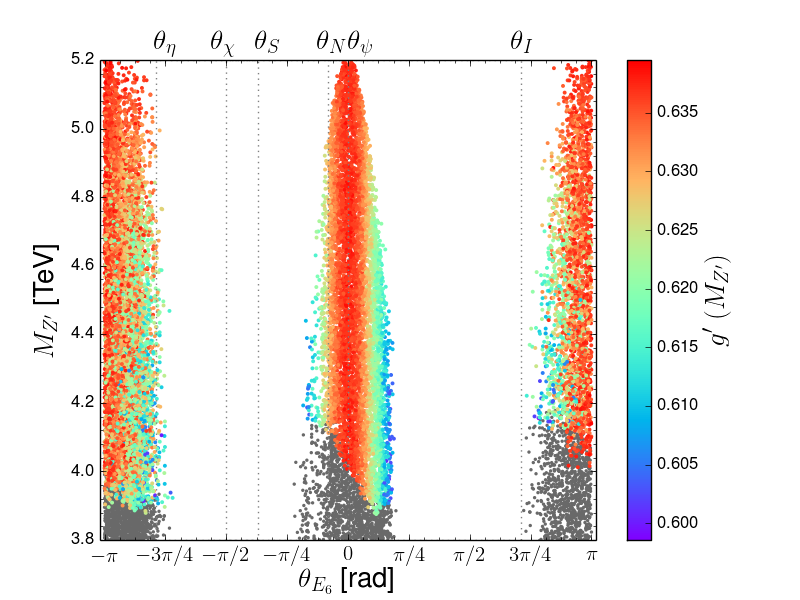}
 \caption{In the upper panel, we compare the $\sigma {\rm B}$ rate with ATLAS
  data,
  regardless of the specific $U(1)^\prime$ group and emphasizing the
  values of the $Z^\prime\to ll$ branching ratio. In the lower panel, we show
  the correlations between the $Z'$-boson mass and the $\theta_{E_6}$
  mixing angle for all points satisfying the constraints detailed in
  Section~\ref{subsec:scan}. Points that are excluded at
  the $2\sigma$ level by the recent ATLAS search for $Z'$ in the dilepton
  mode~\cite{Aaboud:2017buh} are shown in grey, whilst the value of the
  $U(1)^\prime$ coupling strength is shown otherwise. Both figures refer to
  the scenario where couplings unify at $M_{\rm GUT}$.}
 \label{fig:sigBscan1}
\end{figure}

In Fig.~\ref{fig:sigBbench1} we compare the ATLAS limits on high-mass dileptons
at the $1\sigma$ (green) and $2\sigma$ (yellow) levels
with our predictions  for $\sigma{\rm B}$, obtained 
in the context of $U(1)'_\psi$ and $U(1)'_\eta$ 
(upper panel), as well as $U(1)^\prime_I$ and $U(1)^\prime_N$ (lower panel)
gauge groups, in the range \mbox{$2~{\rm TeV}<M_{Z^\prime}<5~{\rm TeV}$}.
We consider both supersymmetric (markers with error bars) and 
non-supersymmetric cases (dashed lines) and include NLO
QCD corrections to the production cross section $\sigma(pp\to Z^\prime)$.
The error bars around the supersymmetric
results include two contributions: first, they  account for
the spread covered in the
scan and second, they include the theoretical uncertainties stemming from traditional
scale and
parton density variations in the NLO computation.
We found that the latter uncertainty
varies from 5\% for $Z^\prime$ masses of about
2~TeV and goes up to 20\% for $M_{Z^\prime}\simeq 5$~TeV.
We observe that the impact of supersymmetric
decays on the excluded $M_{Z^\prime}$ values
runs from about 100 GeV ($Z^\prime_\eta$) to
200 GeV ($Z^\prime_\psi$ and $Z^\prime_N$), while the errors on the
$Z^\prime_I$ dilepton rate in the UMSSM are too large to discriminate it
from the non-supersymmetric case.
Overall, 
$Z^\prime$ bosons lighter than 4 TeV are still strongly disfavored by
ATLAS data, regardless of the $U(1)^\prime$ model.

In Fig.~\ref{fig:sigBscan1} (upper panel), we reexpress the same
results by emphasizing the dependence of $\sigma{\rm B}$ on the
dilepton branching fraction, by
superimposing the predictions of the different $U(1)^\prime$
realizations, regardless of the actual
$\theta_{E_6}$ mixing angle, and displaying the values
of ${\rm BR}(Z^\prime\to ll)$ by means of different colors.
We find that the dilepton rate varies between 4\% and 12\%, 
and that the yielded exclusion masses are roughly between
4 and 4.5 TeV.

\begin{figure}
 \centering
 \includegraphics[width=0.7\columnwidth]{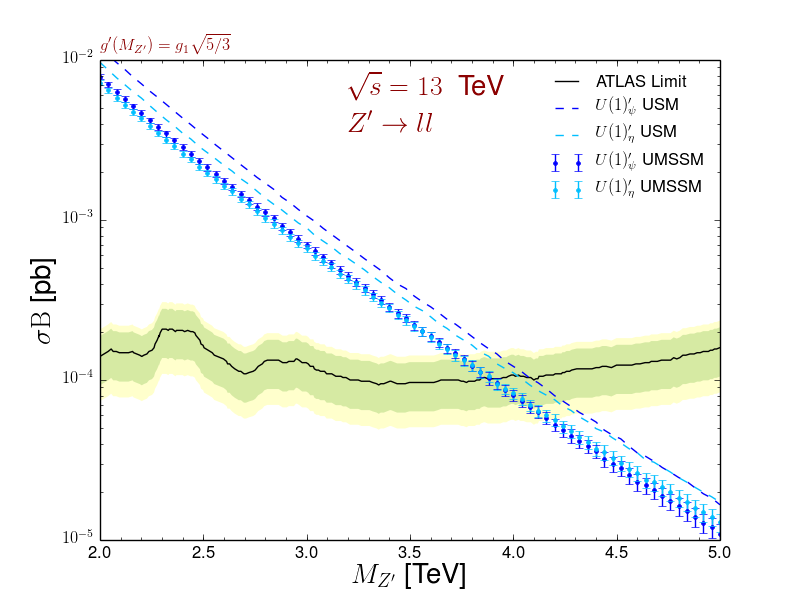}
 \includegraphics[width=0.7\columnwidth]{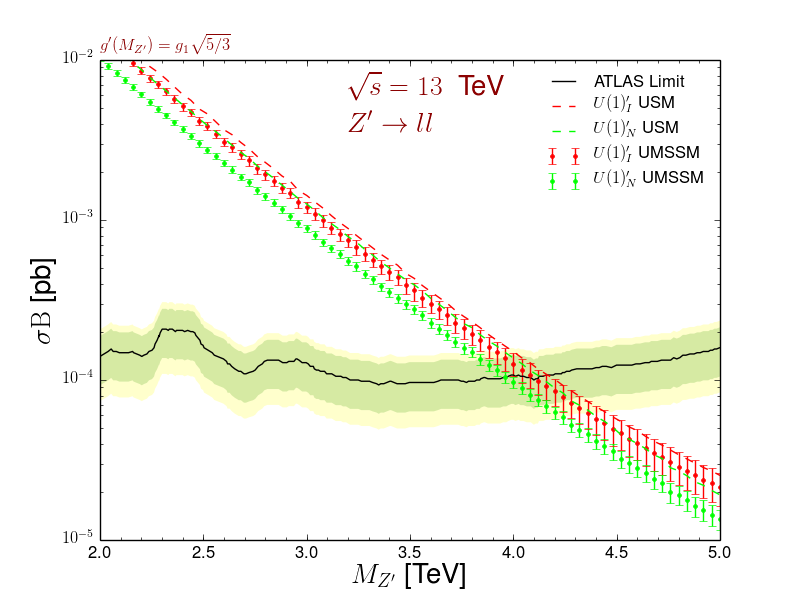}
 \caption{As in Fig.~\ref{fig:sigBbench1}, but for the
  scenario where the condition $g^\prime=\sqrt{5/3}g_1$ is imposed at
  $M_{Z^\prime}$.}
 \label{fig:sigBbench2}
\end{figure}

In the lower panel of Fig.~\ref{fig:sigBscan1}, we present
instead the distribution of
the allowed $Z^\prime$-boson masses
as a function of the $E_6$ mixing angle, with 
the value of the $g^\prime$ coupling for each
scenario indicated by a color code.
In order to determine the allowed regions, we first impose the 
experimental constraints in Table~\ref{tab:constraints} and then
the exclusion limits coming from the direct comparison with the
ATLAS data in Fig.~\ref{fig:sigBbench1}.
The points ruled out by the ATLAS results 
are shown in grey.
We observe, similarly to the findings of
Ref.~\cite{Araz:2017qcs}, that only $|\theta_{E_6}|$ values in the
intervals $[0,\pi/4]$ and $[3/4\pi, \pi]$
can accommodate all the imposed experimental
constraints. Outside of these regions,
the $U(1)^\prime$ charge of the extra singlet
supermultiplet $S$ is in fact
close to zero so that either the SM-like
Higgs boson or the $Z^\prime$ boson, or even
both, are predicted to be too light with respect to current data.
In particular, Fig.~\ref{fig:sigBscan1} (lower panel)
dictates that models $U(1)^\prime_\chi$ and $U(1)^\prime_S$
are largely ruled out by the current data (see also the above discussion), while
$U(1)^\prime_I$ is only marginally consistent.
As a whole, after adding the recent ATLAS constraints \cite{Aaboud:2017buh}
(the grey points), it turns out once again that
scenarios exhibiting a $Z^\prime$ boson
lighter than 4~TeV can hardly ever be realized, the
corresponding parameter-space regions getting more and more restricted.

\subsection{Scenarios with Low-Scale Boundary Conditions}
\label{subsec:bt}
In this subsection, we focus on the second class of scenarios, 
wherein the input
parameters, given in Eq.~\eqref{eq:prms2}, are provided at the
$Z^\prime$ mass scale and where the $U(1)^\prime$ coupling reads
\be
\label{gplow}
g^\prime (M_{Z^\prime})= \sqrt{\frac{5}{3}}~g_1 (M_{Z^\prime}) \approx 0.47\ ,
\ee
for all models satisfying the constraints imposed in
subsection~\ref{subsec:scan}.
Comparing Eq.~(\ref{gplow}) with the minimal values for
$g^\prime(M_{Z^\prime})$
quoted in Table~\ref{tab:sce1assump}, we learn that, for low-scale boundary
conditions, $g^\prime$ is substantially smaller.
Therefore,  the $Z^\prime$-production cross section is 
lower than for scenarios where boundary
conditions are provided at the GUT scale $M_{\rm GUT}$.

\begin{figure}
 \centering
 \includegraphics[width=0.7\columnwidth]{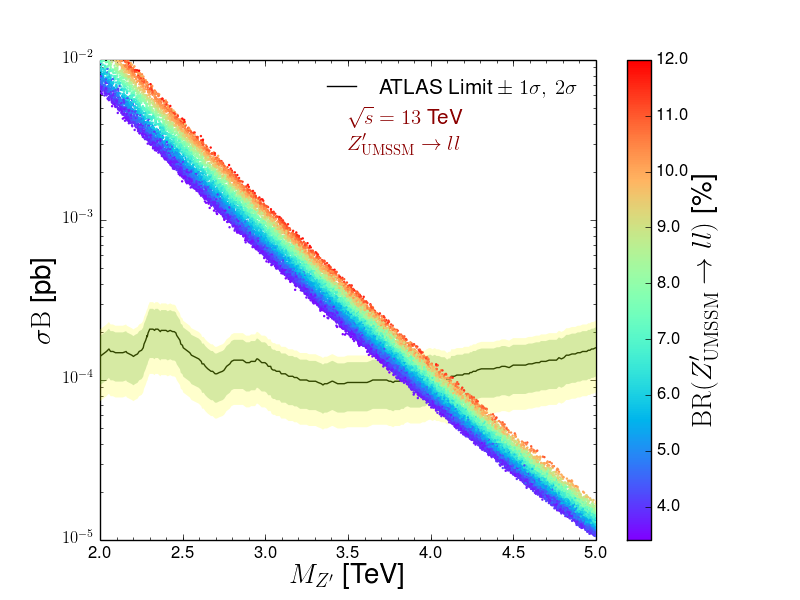}
 \includegraphics[width=0.7\columnwidth]{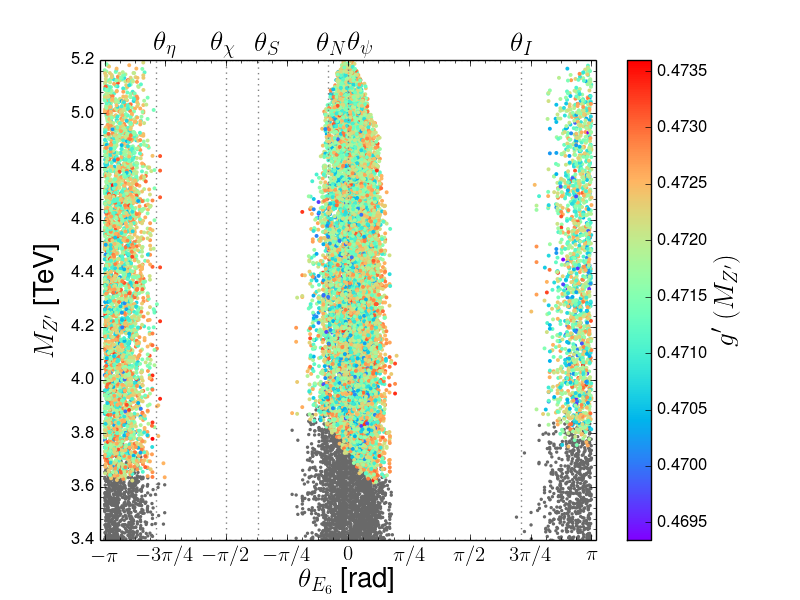}
 \caption{As in Fig.~\ref{fig:sigBscan1}, but for coupling unification
  at $M_{Z^\prime}$.}
 \label{fig:sigBscan2}
\end{figure}

As a consequence, the inferred $Z^\prime$ mass
exclusion limits are 
reduced by about 200--300~GeV with respect to the
high-scale unification case,
as can be seen in Fig.~\ref{fig:sigBbench2}, where the ATLAS
limits are compared with the UMSSM predictions for $U(1)^\prime_\psi$,
$U(1)^\prime_\eta$ (upper panel), and $U(1)^\prime_N$ and $U(1)^\prime_I$ (lower
panel) models.
Since the $g^\prime$ value is roughly the same as in the non-supersymmetric
case, the overall impact of the inclusion of supersymmetric
decays is similar to that found in the high-scale boundary
framework, namely a reduction of the bounds on the $Z'$ boson mass by about
200~GeV.
As observed for the other class of scenarios, the models with the highest impact
of novel decay modes are the $U(1)^\prime_\eta$ and $U(1)^\prime_N$ ones, while
the errors are too large to appreciate the effect of non-standard decays in
$\sigma{\rm B}$ for the $U(1)^\prime_I$ case. Our analysis then confirms
the finding of Ref.~\cite{Corcella:2013cma},
which compared UMSSM predictions in the low-scale
unification framework with 8 TeV LHC limits and 
obtained an effect of similar magnitude on the excluded masses.

As  for the high-scale unification case,
we present in Fig.~\ref{fig:sigBscan2} (upper panel) 
the comparison of $\sigma {\rm B}$ with the ATLAS data, scanning through the
whole parameter space and displaying in different color codes 
the values of ${\rm BR}(Z^\prime\to ll)$. Fig.~\ref{fig:sigBscan2}
(lower panel) shows instead 
the correlations between the allowed $M_{Z^\prime}$ values and 
$\theta_{E_6}$, accounting for both indirect
constraints and direct ATLAS exclusion limits, the latter
given by the grey-shaded area.
The results in Fig.~\ref{fig:sigBscan2} are qualitatively similar to those
presented in Fig.~\ref{fig:sigBscan1}. However, as anticipated before, the
$g^\prime$ value is smaller, so that the ATLAS constraints on $M_{Z^\prime}$ are
milder and values of $M_{Z^\prime}\gsim 3.6$~TeV are hence still allowed.
Likewise, regarding specific $U(1)^\prime$ models,
$U(1)^\prime_\chi$ and $U(1)^\prime_S$ are ruled out,
while the other setups are still permitted and worth to be further explored.

\section{Leptophobic $Z^\prime$ Scenarios in UMSSM Models}
\label{sec:leptophobic}
The results presented in the previous section have shown that the inclusion
of supersymmetric decays has a substantial effect on the $Z^\prime$ searches
and exclusion limits, but nevertheless the
ATLAS bounds
originating from the dilepton channel strongly constrain any
phenomenologically viable UMSSM realization. Furthermore, the very fact that
the $Z^\prime$ boson has to be quite heavy
impacts all sfermion masses through 
the $U(1)^\prime$ $D$-terms, which may even lead
to discarding some scenarios, such as $U(1)^\prime_\chi$, as 
yielding unphysical sfermion spectra.
All LHC constraints studied so far can, however, be evaded
by enforcing the $Z^\prime$ boson to be leptophobic.
In these scenarios, resonance searches in the dijet
final state become the main probes of the new boson, Run~II results for the
top-antitop mode including the analysis of the full 2016 dataset being still
not available. Dijet bounds are however
much weaker, as described in Refs.~\cite{Sirunyan:2016iap,Aaboud:2017yvp}.

\begin{figure}
 \centering
 \includegraphics[width=0.7\columnwidth]{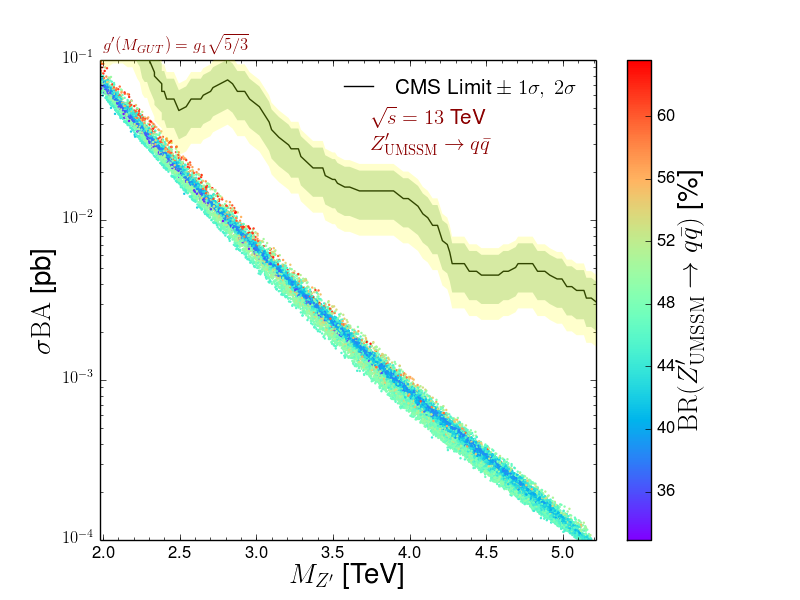}
 \includegraphics[width=0.7\columnwidth]{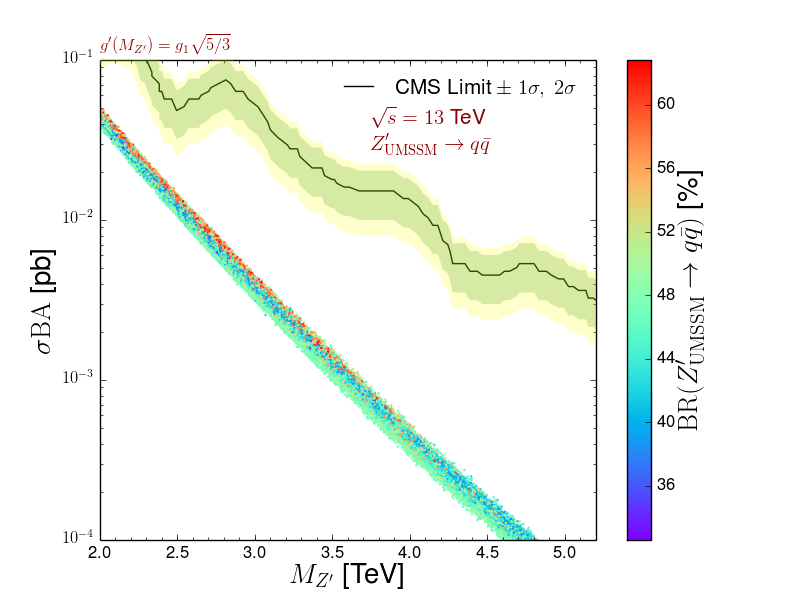}
 \caption{$Z^\prime$ production cross section multiplied by 
  the dijet branching ratio and by the acceptance ${\rm A}\simeq 0.6$,
  for the first (upper panel) and second (lower panel) class of scenarios
  investigated in this work. We compare NLO QCD theoretical predictions
  to the bounds obtained
  by the CMS collaboration~\cite{Sirunyan:2016iap} at the $1\sigma$ (green)
  and $2\sigma$ (yellow) level.
  The actual $Z^\prime$ dijet branching ratio is
  indicated with the color code.}
 \label{fig:Zpqq}
\end{figure}

Before discussing the phenomenology of leptophobic $Z^\prime$ bosons
within supersymmetry, 
in Fig.~\ref{fig:Zpqq} we compare the CMS high-mass
dijet yield from Ref.~\cite{Sirunyan:2016iap} with our predictions for
$\sigma(pp\to Z^\prime)\times{\rm BR}(Z^\prime \to q\bar q)$, obtained after
scanning the UMSSM parameters as described in Table~\ref{tab:scan_lim} and
imposing the constraints of Table~\ref{tab:constraints},
for scenarios with high-scale (upper panel)
and low-scale (lower panel) boundary conditions.
As in the dilepton channel, the production cross section
is calculated at NLO and the values
of the dijet branching
ratios are characterized by different color codes.
For the sake of consistency with the experimental analysis,
the $\sigma{\rm B}$ rate is multiplied by an acceptance factor
${\rm A}\simeq 0.6$ and the fraction of $Z^\prime\to t\bar t$ events
is not included in the calculation.

From Fig.~\ref{fig:Zpqq}, one learns that the computed $\sigma {\rm B}{\rm A}$
is always below the CMS exclusion limits in the range
$2~{\rm TeV}<M_{Z^\prime}<5~{\rm TeV}$ at the 95\% confidence level
in both frameworks of coupling unification, 
once accounting for supersymmetric $Z^\prime$ decays.
One can, therefore, envisage than even much 
lighter $Z'$ bosons could be allowed by data when leptophobic 
UMSSM realizations,
such as those introduced in Section~\ref{subsec:u1prime}, are
considered.

\begin{figure}
 \begin{center}
  \includegraphics[width=0.7\columnwidth]{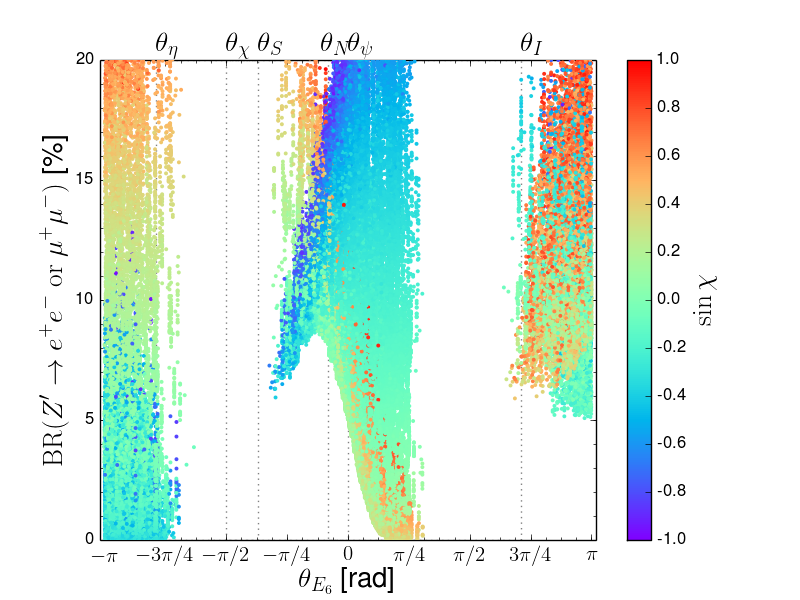}
 \end{center}
 \caption{Correlations between the $Z^\prime$-boson
  branching ratio into a dilepton
  system and the $\theta_{E_6}$ mixing angle featured by all points satisfying
  the constraints detailed in Section~\ref{subsec:scan} and for UMSSM
  scenarios where the input parameters are fixed at the $Z'$ mass scale
  (second class of considered scenarios). The value of the
  sine of the kinetic mixing angle ($\sin \chi$) is indicated by the color code.}
 \label{fig:Qbarscan}
\end{figure}

Hereafter we focus on the second class of UMSSM scenarios,
{\it i.e.}, coupling unification at the $M_{Z^\prime}$ scale, 
and add to the list of free parameters in Eq.~\eqref{eq:prms2} the
sine of the kinetic
mixing angle $\sin\chi$, defined through Eq.~(\ref{bbzz}), 
that we allow to vary in the $[-1, 1]$ window.
In principle, as thoroughly debated in Ref.~\cite{Belanger:2017vpq}, the kinetic
mixing angle also affects the Dark Matter relic abundance, since
the mass of the lightest supersymmetric particle (LSP),
which in Ref.~\cite{Belanger:2017vpq} can be either a
right-handed neutrino or the lightest neutralino and in this
paper is $\tilde\chi^0_1$, 
depends on the $U(1)^\prime$ charges of the Higgs bosons, which
have been modified according to Eq.~(\ref{eq:Qbardef}) and 
are a function of $\sin\chi$\footnote{Note that the kinetic mixing
  parameter $k$ in Ref.~\cite{Belanger:2017vpq} corresponds
  to our $\sin\chi$.}.
Because of that, the authors of Ref.~\cite{Belanger:2017vpq},
besides applying the constraints due to collider physics, 
accounted for the upper
bound on the relic density as well, relying on the 
Planck 2015 measurements \cite{Ade:2015xua}.
The finding of Ref.~\cite{Belanger:2017vpq} is that,
although the mass of the LSP is indeed sensitive
to $\sin\chi$ and, e.g., a heavy Dark Matter candidate is favored by
small $|\sin\chi|$, 
a value of the relic density $\Omega h^2\sim 0.1$,
consistent with Ref.~\cite{Ade:2015xua}, 
can be achieved for any value of $\sin\chi$, and 
in particular for $|\sin\chi|\simeq 0.3$, corresponding to a leptophobic
$Z'$.
In view of these results, we shall
not impose further constraints, beyond those already discussed in the
previous sections, and assume that any $\sin\chi$ can possibly be consistent
with the Dark Matter relic density, including the values which make
the $Z'$ leptophobic.

In Fig.~\ref{fig:Qbarscan}, we present the $Z^\prime$
dilepton branching ratio, scanning the parameter space as presented in
Section~\ref{subsec:scan}, in terms of the mixing angle 
$\theta_{E_6}$ and $\sin\chi$.
In agreement with Eq.~\eqref{eq:Qbardef}, we realize that 
values of $\sin\chi$ around $\pm0.3$ can lead to leptophobia whenever 
the $E_6$ mixing angle obeys the condition in 
Eq.~\eqref{eq:e6mixlept} and the $U(1)^\prime$
charges fulfill the relation $\bar{Q}_E \approx \bar{Q}_L \approx 0$.
In particular, the condition ${\rm BR}(Z^\prime\to l^+l^-)\simeq 0$
can be achieved for $-\pi\lsim\theta_{E_6}\lsim -3\pi/4$, which
includes the $U(1)^\prime_\eta$ model, and for
$\pi/8\lsim\theta_{E_6}\lsim \pi/4$, hence in the neighborhood of
$U(1)^\prime_\psi$. The other $U(1)^\prime$ symmetries are either
ruled by the experimental data or, even in the
most optimistic case, can hardly lead to dilepton rates
below 5\%.

Of course, these leptophobic
scenarios cannot be constrained by standard $Z^\prime$-boson
searches in dimuons or dielectrons 
at the LHC, and novel strategies must be designed.
In the following, we propose a
selection potentially allowing to observe leptophobic light
$Z^\prime$ bosons decaying through a supersymmetric cascade.
As direct decays are forbidden, 
dilepton final states can arise from ($Z^\prime$-mediated)
chargino-pair production and subsequent decays into a charged lepton
and missing energy via an intermediate $W$ boson, possibly off-shell,
namely
$\tilde \chi_1^\pm \to (W^\pm \to l^\pm \nu_l)\  \tilde\chi^0_1$,
$\tilde\chi^0_1$ being the lightest neutralino.
However, for the points selected by our scan procedure, the off-shell
contributions are typically either negligible (when the two-body decay channel
is open) or not important enough to yield a sufficient number of signal events
(when the $\tilde\chi^\pm_1\to W^\pm \tilde\chi^0_1$ decay is closed).
We, therefore, design an analysis assuming the presence of intermediate
on-shell $W$ bosons, targeting thus UMSSM scenarios where the mass
difference between the lightest chargino $\tilde\chi^\pm_1$ and
the lightest neutralino $\tilde\chi^0_1$ is at least $M_W\simeq 80$~GeV.
The signal process
consists of the resonant production of a chargino pair, followed by the
decay of each chargino into a charged lepton and missing energy,
\be\label{chain}
pp \to Z^\prime \to \tilde \chi^+_1 \tilde \chi^-_1 \to l^+l^- + \slashed{E}_{T} \ .
\ee

We focus on two optimistic signal benchmarks that
are currently not excluded by data and with different $U(1)^\prime$
properties. Both scenarios exhibit 
a $Z^\prime$ boson with a
mass of about 2.5 TeV and charginos and neutralinos as light as possible,
in order to maximize the branching ratios in Eq.~(\ref{chain}), 
but with a mass splitting larger than $M_W$,
in such a way to
allow the transition
$\tilde\chi^\pm_1\to\tilde\chi^0_1 W^\pm$ with real $W$ bosons.
The
first scenario, that we denote {\bf BM I},
relies on a $U(1)^\prime_\eta$
symmetry, namely $\theta_{E_6} = -0.79\pi$, since UMSSM
scenarios based on this specific gauge symmetry can be made
naturally leptophobic, as shown in Fig.~\ref{fig:Qbarscan}.
The second scenario, dubbed {\bf BM II},
has instead a symmetry close to the
$U(1)^\prime_\psi$ setup, but with a larger
mixing angle, {\it i.e.} $\theta_{E_6} = 0.2\pi$, so that a
leptophobic $Z^\prime$
boson can still be realized (see again Fig.~\ref{fig:Qbarscan}).

\begin{table}
 \renewcommand{\arraystretch}{1.4}
 \begin{center}
  \begin{tabular}{c||c|c|c|c|c|c}
   Parameter &
     $\theta_{E_6}$ & $\tan\beta$ & $\mu_{\rm eff}$ [GeV] &
     $M_{Z^\prime}$ [TeV] & $M_0$ [TeV] &  $M_1$ [GeV]\\ \hline \hline
   {\bf BM I} &
     $-0.79~\pi$ & 9.11 & 218.9  & 2.5 & 2.6 & 106.5 \\ \hline
   {\bf BM II} &
     $0.2~\pi$  & 16.08 & 345.3 & 2.5 & 1.9 & 186.7\\[.4cm]
  \multicolumn{7}{c}{}\\
   Parameter &
     $M_2$ [GeV] & $M_3$ [TeV] & $M_1^\prime$ [GeV] & $A_0$ [TeV] & 
     $A_\lambda$ [TeV]& $\sin\chi$ \\ \hline \hline
   {\bf BM I} &
     230.0 & 3.6 & 198.9 & 2 & 5.9 & $-0.35$ \\ \hline
   {\bf BM II} &
     545.5 & 5.5 & 551.7 & 1.5 & 5.1 & 0.33
  \end{tabular}
  \caption{\label{parabm}
   UMSSM parameters for the reference points
   {\bf BM I} and {\bf BM II}.}\end{center}
\end{table}

The UMSSM parameters for the two points are quoted in
Table~\ref{parabm}, while Tables~\ref{tabmass1} and
\ref{tabmass2} contain the predicted masses for
gluinos, squarks, sleptons, Higgses and gauginos in 
the reference points {\bf BM I} and {\bf BM II}, respectively.
The branching ratios of the $Z^\prime$ in such representative points
are listed in Table~\ref{tabbr1}, omitting rates which are below 1\%.

\begin{table}
 \renewcommand{\arraystretch}{1.3}
 \begin{center}
  \begin{tabular}{|c|c|c|c|c|c|c|}
   \hline
   $M_{\tilde g}$ & $M_{\tilde d_1}$ &   $M_{\tilde u_1}$ &  $M_{\tilde s_1}$ &  $M_{\tilde c_1}$ &  
   $M_{\tilde b_1}$ & $M_{\tilde t_1}$\\
   \hline
   3745.1 & 2988.8 & 2937.3  & 3380.3 & 3025.9 & 3380.4 & 3379.4 \\
   \hline\hline
   & $M_{\tilde d_2}$ &   $M_{\tilde u_2}$ &  $M_{\tilde s_2}$ &  $M_{\tilde c_2}$ &  
   $M_{\tilde b_2}$ & $M_{\tilde t_2}$ \\ 
   \hline
   & 3525.2 & 3379.4 &  3541.2 & 3699.0 & 3541.2 & 3699.0 \\
   \hline\hline
   & $M_{\tilde e_1}$ &   $M_{\tilde e_2}$ & $M_{\tilde \mu_1}$ & $M_{\tilde \mu_2}$ &
   $M_{\tilde\tau_1}$ & $M_{\tilde\tau_2}$  \\ 
   \hline
   & 171.1 & 345.7  & 196.4 & 392.3 & 239.4 & 409.6  \\
   \hline\hline
   & $M_{\tilde \nu_{e,1}}$ &  $M_{\tilde \nu_{e,2}}$ &
   $M_{\tilde \nu_{\mu,1}}$ &  $M_{\tilde \nu_{\mu,2}}$ &
   $M_{\tilde \nu_{\tau,1}}$ &  $M_{\tilde \nu_{\tau,2}}$ \\
   \hline
   & 336.4 & 1663.1 & 384.1&
   1674.2 & 401.6 & 1683.6 \\
   \hline\hline
   $M_h$ &   $M_H$&  $M_{H^\prime}$ &  $M_A$ & $M_{H^\pm}$ &
   $M_{\tilde\chi_1^+}$ &   $M_{\tilde\chi_2^+}$ \\
   \hline
   122.5 &  3371.5 & 2507.0  & 3371.5 & 3372.7 &
   177.1 & 302.3 \\
   \hline\hline
   & $M_{\tilde\chi_1^0}$ &   $M_{\tilde\chi_2^0}$ 
   & $M_{\tilde\chi_3^0}$ &   $M_{\tilde\chi_4^0}$ & $M_{\tilde\chi_5^0}$ &   $M_{\tilde\chi_6^0}$\\
   \hline
   & 95.5  & 181.3 & 232.2 & 302.4 & 2405.1 & 2602.0\\
   \hline\end{tabular}
  \caption{Masses of gluino, squarks, sleptons, Higgs and gauginos 
   for the UMSSM benchmark point {\bf BM I}.
   $\tilde q_{1,2}$, $\tilde l_{1,2}$ and
   $\tilde\nu_{1,2}$ are mass eigenstates and 
   differ from the gauge eigenstates
   $\tilde q_{L,R}$,  $\tilde\ell_{L,R}$ and $\tilde\nu_{L,R}$
   by virtue of the mass mixing contributions that are relevant especially in
   the stop case. All masses are in GeV.\label{tabmass1}}\vspace{.8cm}
  \begin{tabular}{|c|c|c|c|c|c|c|}
   \hline
   $M_{\tilde g}$ & $M_{\tilde d_1}$ &   $M_{\tilde u_1}$ &  $M_{\tilde s_1}$ &  $M_{\tilde c_1}$ &  
   $M_{\tilde b_1}$ & $M_{\tilde t_1}$\\ 
   \hline
   5669.3 & 4405.5 & 4141.5 & 4927.6 & 4418.1 & 4927.7 & 4926.9 \\
   \hline\hline
   & $M_{\tilde d_2}$ &   $M_{\tilde u_2}$ &  $M_{\tilde s_2}$ &  $M_{\tilde c_2}$ &  
   $M_{\tilde b_2}$ & $M_{\tilde t_2}$ \\ 
   \hline
   &
   5069.8 & 4927.0 & 5146.3 & 5117.1 & 5146.3 & 5117.1\\
   \hline\hline
   & $M_{\tilde e_1}$ &   $M_{\tilde e_2}$ & $M_{\tilde \mu_1}$ & $M_{\tilde \mu_2}$ &
   $M_{\tilde\tau_1}$ & $M_{\tilde\tau_2}$ \\ 
   \hline
   & 665.1 & 871.5  & 679.2 & 1067.9 & 743.9 & 1075.6 \\
   \hline\hline
   &
   $M_{\tilde \nu_{e,1}}$ &  $M_{\tilde \nu_{e,2}}$ &
   $M_{\tilde \nu_{\mu,1}}$ &  $M_{\tilde \nu_{\mu,2}}$ &
   $M_{\tilde \nu_{\tau,1}}$ &  $M_{\tilde \nu_{\tau,2}}$ \\ 
   \hline
   & 660.4  & 1049.6 & 674.3 &1079.4 & 739.3  & 1106.2 \\
   \hline\hline
   $M_h$ &   $M_H$&  $M_{H^\prime}$ &  $M_A$ & $M_{H^\pm}$ &
   $M_{\tilde\chi_1^+}$ &   $M_{\tilde\chi_2^+}$ \\
   \hline
   127.4 &   5237.8  &   2498.2  & 5238.0 & 5238.8 &  343.8 & 593.5 \\
   \hline\hline
   & $M_{\tilde\chi_1^0}$ &   $M_{\tilde\chi_2^0}$ 
   & $M_{\tilde\chi_3^0}$ &   $M_{\tilde\chi_4^0}$ & $M_{\tilde\chi_5^0}$ &   $M_{\tilde\chi_6^0}$\\
   \hline
   & 178.1  &346.9 & 360.0 & 593.2 & 2239.1 & 2785.9\\
   \hline\end{tabular}
  \caption{Same Table~\ref{tabmass1} but for the UMSSM benchmark point
   {\bf BM II}.
   \label{tabmass2}}
 \end{center}
\end{table}

Table~\ref{parabm} shows that {\bf BM II} features substantially
larger values of
$\tan\beta$, $\mu_{\rm eff}$ and of the gaugino masses $M_1$, $M_2$, $M_3$ and
$M_4$, while $M_0$ and the trilinear couplings $A_0$ and $A_\lambda$ are
smaller than in {\bf BM I}.
Comparing Tables~\ref{tabmass1} and \ref{tabmass2}, one learns that
in {\bf BM I} the squarks have masses between 3 and 4 TeV, while
in {\bf BM II} they are on average more than 1 TeV heavier.
Charged sleptons in {\bf BM II} are instead lighter than in {\bf BM I},
unlike sneutrinos, whose masses vary between about 300 GeV and 1.7 TeV
in {\bf BM I} and between 660 GeV and 1.1 TeV 
in {\bf BM II}. 
In the Higgs sector, with the exception of the SM-like $h$, all
Higgs bosons have masses of a few TeV and are therefore too heavy
to contribute to $Z^\prime$ decays for both benchmarks.

In particular, as anticipated,
the singlet-like neutral boson $H^\prime$ has approximately
the same mass as the $Z^\prime$, while $H$, $A$ and $H^\pm$ are
roughly degenerate, with mass about 3.37 TeV in {\bf BM I}
and 5.24 TeV in {\bf BM II}.
As for gauginos, as anticipated, the two novel neutralinos
$\tilde\chi^0_5$ and $\tilde\chi^0_6$ have masses similar to $M_{Z^\prime}$,
thus too high to be relevant
for $Z^\prime$ decays, while charginos and MSSM-like neutralinos
are sufficiently light to possibly contribute to the $Z^\prime$ width.
Overall, the electroweakino spectrum is more compressed in the
reference point {\bf BM I}.
The mass splitting between $\tilde\chi^\pm_1$ and $\tilde\chi^0_1$ is
in fact slightly above $M_W$ in {\bf BM I}, while it is substantially larger than $M_W$, {\it i.e.} about 165 GeV, in the {\bf BM II} framework.
In both cases, the decay $\tilde\chi^\pm_1\to W^\pm\tilde\chi^0_1$
can occur through on-shell $W$-bosons and has a branching fraction of 
almost 100\%.

Concerning the $Z^\prime$ branching ratios, Table~\ref{tabbr1}  shows that the
branching fraction of the $Z^\prime$ boson decay into a
$\tilde\chi^+_1\tilde\chi^-_1$ pair, entering in the process of
Eq.~\eqref{chain}, is of about 2\% for the scenario {\bf BM I} and 6\% for the
scenario {\bf BM II}.  {\bf BM I} allows for substantial branching fractions into other combinations of chargino pairs, while 
both scenarios exhibit non-negligible rates into neutralino pairs, and the
{\bf BM II} scenario also includes decays into sneutrino pairs as well.
The decay rates in pairs of the lightest neutralinos, possible candidates for Dark Matter, are instead suppressed
in both reference points.
As a whole, supersymmetric decays are responsible for 12\% and
15\% of the $Z^\prime$ width in the representative points
{\bf BM I} and {\bf BM II}, respectively.

\begin{table}
 \renewcommand{\arraystretch}{1.4}
 \begin{center}
  \begin{tabular}{c||c | c}
   Decay mode & BR [\%] ({\bf BM I}) &  BR [\%] ({\bf BM II})\\
   \hline\hline
   $Z'\to\tilde\chi_1^+\tilde\chi_1^-$            & 1.7 & 6.3\\ \hline
   $Z'\to\tilde\chi_2^+\tilde\chi_2^-$            & 2.1 &  - \\ \hline
   $Z'\to\tilde\chi_1^\pm\tilde\chi_2^\mp$        & 3.9 &  - \\ \hline
   $Z'\to\tilde\chi_2^0\tilde\chi_2^0$      &  -  & 1.5\\ \hline
   $Z'\to\tilde\chi_2^0\tilde\chi_3^0$      & 1.7 & 3.3\\ \hline
   $Z'\to\tilde\chi_3^0\tilde\chi_3^0$      & 1.9 & 1.9\\ \hline
   $Z'\to\tilde\chi_3^0\tilde\chi_4^0$      & 2.2 &  - \\ \hline
   $Z'\to\sum_i\tilde\nu_i\tilde\nu^\dag_i$ &  -  & 1.6\\ \hline
   $Z'\to hZ$                               & 1.9 & 1.9\\ \hline
   $Z'\to W^+W^-$                           & 3.6 & 3.8\\ \hline
   $Z'\to\sum_i d_i\bar d_i$                & 15.8&14.8\\ \hline
   $Z'\to\sum_i u_i\bar u_i$                & 39.8&40.0\\ \hline
   $Z'\to\sum_i \nu_i\bar \nu_i$            & 23.4&22.8
  \end{tabular}
  \caption{$Z^\prime$ decay rates for the benchmark points {\bf BM I} (second
    column) and {\bf BM II} (third column). Branching ratios below 1\% are
    omitted.
  \label{tabbr1}}\end{center}
\end{table}

Once our representative configurations are set, we carry out a
full Monte Carlo event simulation at the LHC, for a center-of-mass energy 
$\sqrt{s}=14$~TeV. 
Hard-scattering signal
events are generated with \textsc{MadGraph5\_aMC@NLO}, the matrix elements being
convoluted with the NLO set of NNPDF 2.3 parton densities.
The production cross section is then
$\sigma(pp\to Z^\prime)\simeq 120$~fb for both benchmarks.
Parton showers and hadronization are simulated by means of the
{\sc Pythia}~8 program \cite{Sjostrand:2014zea}
(version 8.2.19), and the
response of a typical LHC detector is modelled 
with the {\sc Delphes}~3 package
\cite{deFavereau:2013fsa} (version 3.3.2), employing the \textsc{Snowmass}
parameterization~\cite{Anderson:2013kxz,%
 Avetisyan:2013onh}. The resulting detector-level jets are reconstructed
following the anti-$k_T$ algorithm~\cite{Cacciari:2008gp}
with a radius parameter $R=0.6$, as implemented in
the {\sc FastJet} program (version 3.1.3) \cite{Cacciari:2011ma}.
Moreover, we
consider an average number of pile-up events of 140 and normalize our results to
an integrated luminosity of 3000~fb$^{-1}$.

Regarding the backgrounds, we  consider all processes
leading to final states with two charged leptons and missing energy,
such as vector-boson pairs $VV$, with $V$ being a $W$-boson or a $Z$ boson
decaying leptonically.
However, for the purpose of mimicking an actual experimental analysis,
we account for processes yielding also jets which do not pass the acceptance
cuts. Moreover, since our event simulation includes
hadronization effects, 
we explore the possibility that background leptons
originate from hadron decays as well.
Overall, our backgrounds consist of single vector bosons
($V$) or vector-boson pairs ($VV$), possibly accompanied by jets, as well
as $t\bar t$ and single-top events.
In principle, even direct
chargino production ($pp\to \chi^+_1 \chi^-_1 \to l^+l^- + \slashed{E}_{T}$)
should be considered as a background to the
supersymmetric $Z^\prime$ decays. Nevertheless, as pointed out in
Ref.~\cite{Corcella:2014lha}, the leptons produced in processes with
direct charginos, unlike those coming from $Z^\prime$ events,
are typically pretty soft or collinear to the beams. It is therefore
quite easy to suppress the $pp\to\tilde\chi^+_1 \tilde\chi^-_1$ background
by setting suitable cuts on the lepton transverse momenta.

Lepton and jet candidates that are considered throughout our analysis must have 
transverse momenta $p_T^l$ and $p_T^j$ and pseudorapidities $\eta^l$
and $\eta^j$ satisfying
\be\bsp
&p_T^l \geq 20~{\rm GeV} \qquad\text{and}\qquad |\eta^l| < 1.5 \ , \\
&p_T^j \geq 40~{\rm GeV} \qquad\text{and}\qquad |\eta^j| < 2.4 \ .
\esp\label{eq:presel}\ee
Moreover, in our selection strategy,
we reject lepton candidates that are not at an invariant
angular distance, in the
transverse plane, of at least 0.4 from a jet,
\be
\Delta R(j,l) > 0.4  \ ,
\ee
and only focus on muons that are cleaner objects than electrons, in particular
for the pseudorapidity region considered in Eq.~\eqref{eq:presel}. We finally
enforce the considered muons to be isolated, so that the activity in a cone of
radius $R=0.4$ centered on each muon contains at most 15\% of the muon $p_T$,
\be
  I_{\rm rel}^\mu < 0.15 \ .
\ee

We select events featuring two well-separated muons,
since the two signal leptons
$l_1$ and $l_2$ are expected to originate from two different supersymmetric
cascade decays, by requiring
\be
N^l = 2 \qquad\text{and}\qquad \Delta R(l_1,l_2) > 2.5
\ee
and we veto the presence of jets, {\it i.e.}
\be
N^j = 0 \ .
\ee
Furthermore,
the two signal leptons are expected to be produced from the decay of a
heavy $Z^\prime$ with a mass well above the TeV scale. We consequently impose the
transverse momenta of the two leptons to fulfill
\be
p_T(l_1) > 300~{\rm GeV} \qquad\text{and}\qquad p_T(l_2) > 200~{\rm GeV,}
\ee 
which are very efficient cuts to reduce the remaining SM background.
We finally improve the sensitivity by requiring a large amount of missing
energy,
\be
\slashed{E}_T > 100~{\rm GeV,}
\ee
as could be expected for a signal topology where several neutrinos and
neutralinos escape the detector invisibly.

\begin{table}
 \renewcommand{\arraystretch}{1.4}
 \begin{center}
  \begin{tabular}{c l||c||c|c}
   Step&Requirements & Background  & { \bf BM I} & {\bf BM II}  \\\hline \hline
   0& Initial       & $1.7\times10^{11}$ & $8.8\times10^3$ & $1.9\times10^4$ \\
   1& $N^l=2$       & $6.1\times10^8$   & 401              & 860 \\
   2& Electron veto & $2.9\times10^8$    & 100              & 230\\
   3& $ |\eta^l| <1.5$ &    $1.7\times10^8$    &    76    &    170    \\
   4& $ I^\mu_{\rm rel} <0.15$ &    $7.9\times10^5$    &    63    &    130    \\
   5& $\Delta R(l_1,l_2)>2.5$ & $7.9\times10^5$ & 62      & 130\\
   6& Jet veto      & $7.7\times10^4$    & 57              & 120\\
   7& $p_T(l_1)>300$ GeV & 44 & 36 & 71\\
   8& $p_T(l_2)>200$ GeV & 20 & 19 & 32\\
   9& $\slashed{E}_T > 100$ GeV & 10 & 14 & 27\\ \hline
   & \multicolumn{2}{c||}{$s$}   & $3.77\sigma$ & $7.14\sigma$\\
   & \multicolumn{2}{c||}{$Z_A$} & $3.03\sigma$ & $5.05\sigma$\\
  \end{tabular}
  \caption{Selection strategy aiming at observing a leptophobic UMSSM
   $Z^\prime$ boson decaying into a 
   supersymmetric cascade. For each cut, we provide the expected number of
   surviving events for 3000~fb$^{-1}$ of $pp$ collisions at 
   $\sqrt{s}=14$~TeV for both background and signal 
   benchmark scenarios {\bf BM~I} and {\bf BM~II}. We also quote the
   corresponding significances $s$ and $Z_A$, as defined in Eq.~\eqref{eq:sigs}, with 20\% uncertainity.}
  \label{tab:cutflow}
 \end{center}
\end{table}

The corresponding cutflows are shown in Table~\ref{tab:cutflow}, which
illustrates that, for the two benchmark scenarios under consideration,
background
rejection is sufficiently important for observing the signal despite the low
selection efficiencies. For other possible benchmark choices (not
considered in this work) featuring a heavier $Z^\prime$,
the smaller
production total rate is expected to be compensated by a larger
efficiency of the two selection cuts restricting the transverse momenta of the
two selected leptons.

Denoting the number of selected signal and background events by $S$ and
$B\pm\sigma_B$, we make use of two
standard criteria, labelled as $s$ and $Z_A$, 
to define the LHC sensitivity to the leptophobic $Z^\prime$-boson
signal,
\be\bsp
s  =&\ \frac{S}{\sqrt{B+\sigma_B^2}}\ ,\\
Z_A=&\ \sqrt{ 2\left\{
 (S+B)\ln\left[\frac{(S+B)(S+\sigma^2_B)}{B^2+(S+B)\sigma^2_B}\right] -
 \frac{B^2}{\sigma^2_B}\ln\left[1+\frac{\sigma^2_BS}{B(B+\sigma^2_B)}\right]
 \right\}} \ .
\esp\label{eq:sigs}\ee
In Eq.~(\ref{eq:sigs}), $s$ is the significance
as defined by the CMS Collaboration in Ref.~\cite{Ball:2007zza}\footnote{Following Ref.~\cite{Lista:2017jsy}, the denominator of
$s$ sums in quadrature the intrinsic statistical fluctuation of the
background $\sqrt{B}$ and the uncertainty in the background
$\sigma_B$, thus obtaining
$s=S/\sqrt{(\sqrt{B})^2+\sigma_B^2}$,
leading to (\ref{eq:sigs}).}, whereas
the second method ($Z_A$) is
known to be more suitable (and conservative) when the number of background
events is small~\cite{Cowan:2010js}. The conclusions are however very similar in
both cases, as can be seen
from Table~\ref{tab:cutflow}. For both significance definitions, 
we indeed find that the more compressed scenario {\bf BM~I}
could lead to hints visible at the $3\sigma$ level, whilst the second scenario
{\bf BM~II} is in principle observable at even
more than $5\sigma$. The largest LHC
sensitivity to the latter scenario has a twofold origin. First, the $Z^\prime$-induced
chargino-pair production
cross section is larger by virtue of a greater
${\rm BR}(Z^\prime\to \tilde\chi^+_1\chi^-_1)$ branching ratio.
Second, the heavier chargino mass typically induces harder leptons,
the corresponding selection cuts being thus more efficient.

\begin{figure}
 \includegraphics[scale=0.4]{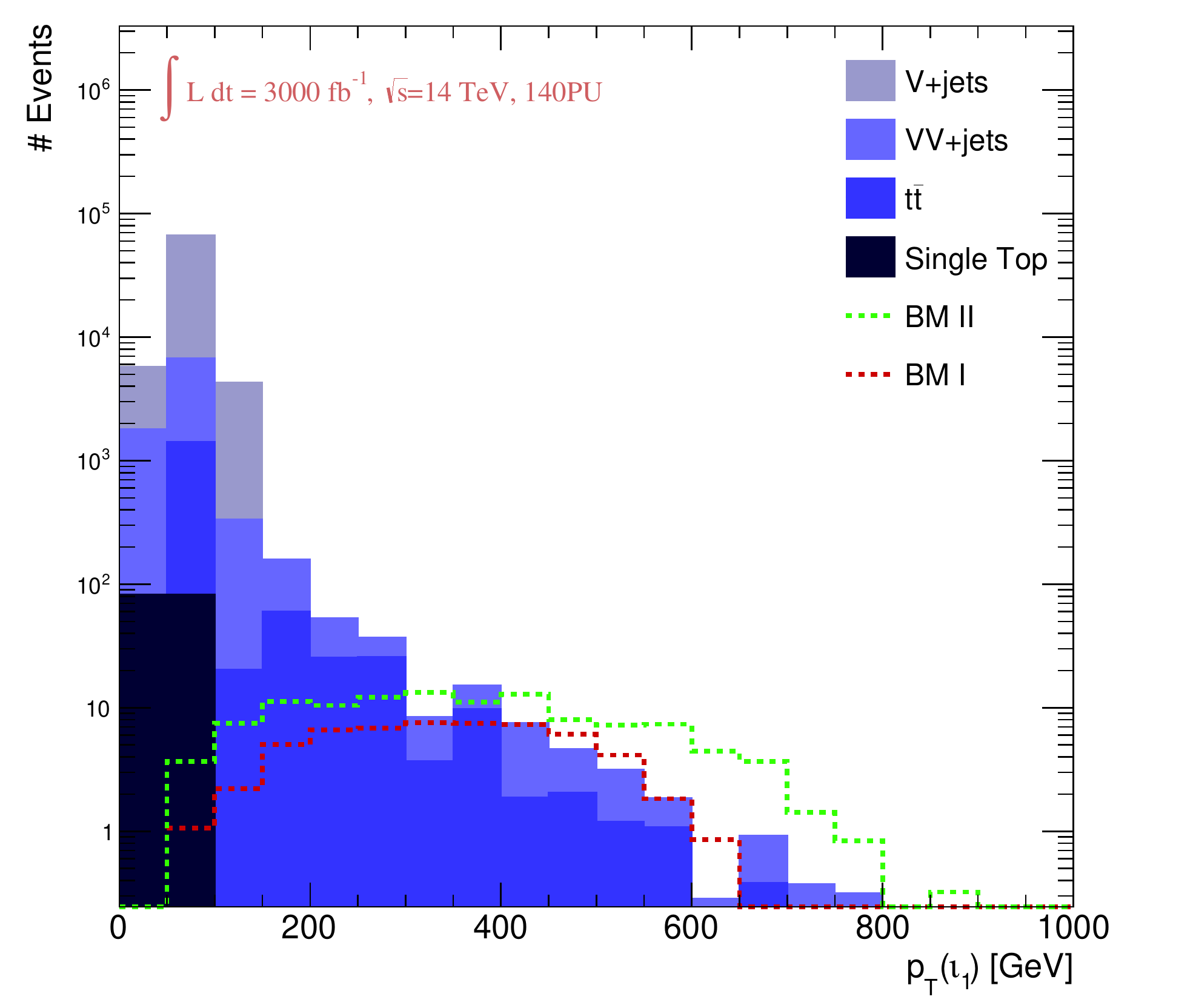}
 \includegraphics[scale=0.4]{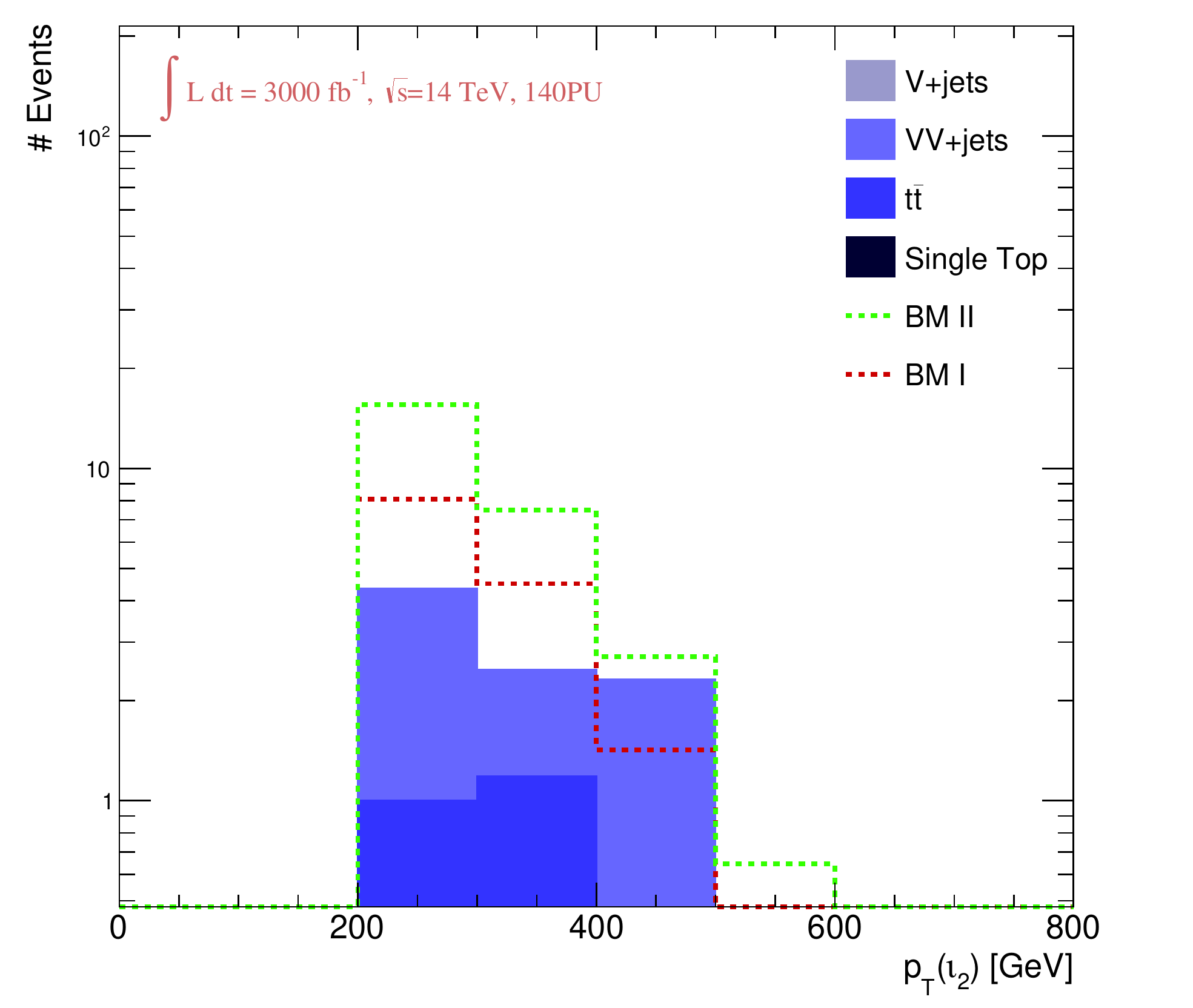}
 \caption{Transverse momentum distribution of the leading muon $l_1$ after
  applying the first 6 cuts of Table~\ref{tab:cutflow} (left) and of the
  next-to-leading muon $l_2$ after applying all cuts (right) for both signal
  scenarios and the backgrounds.}
 \label{fig:pt}
\end{figure}
\begin{figure}
 \includegraphics[scale=0.4]{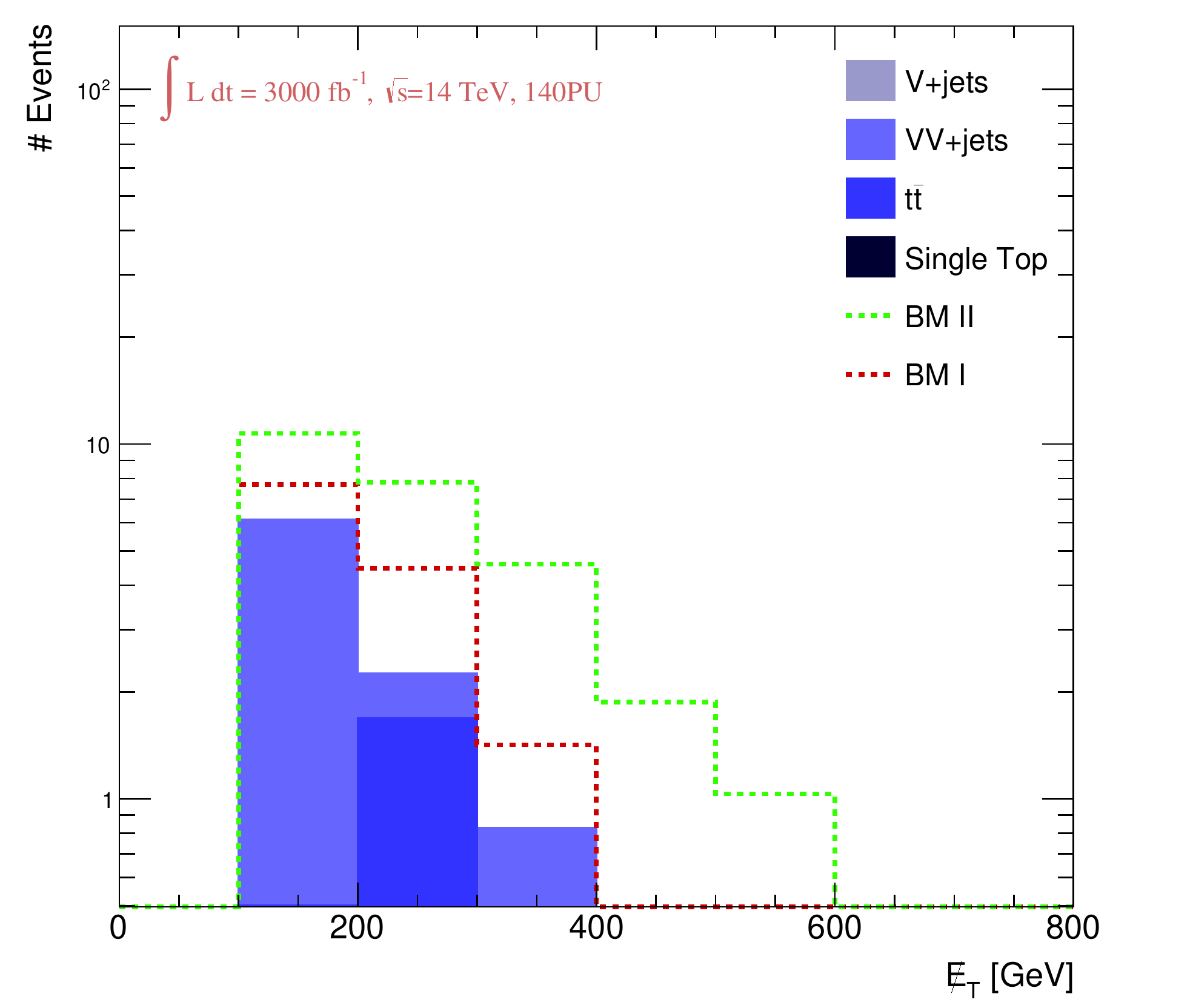}
 \includegraphics[scale=0.4]{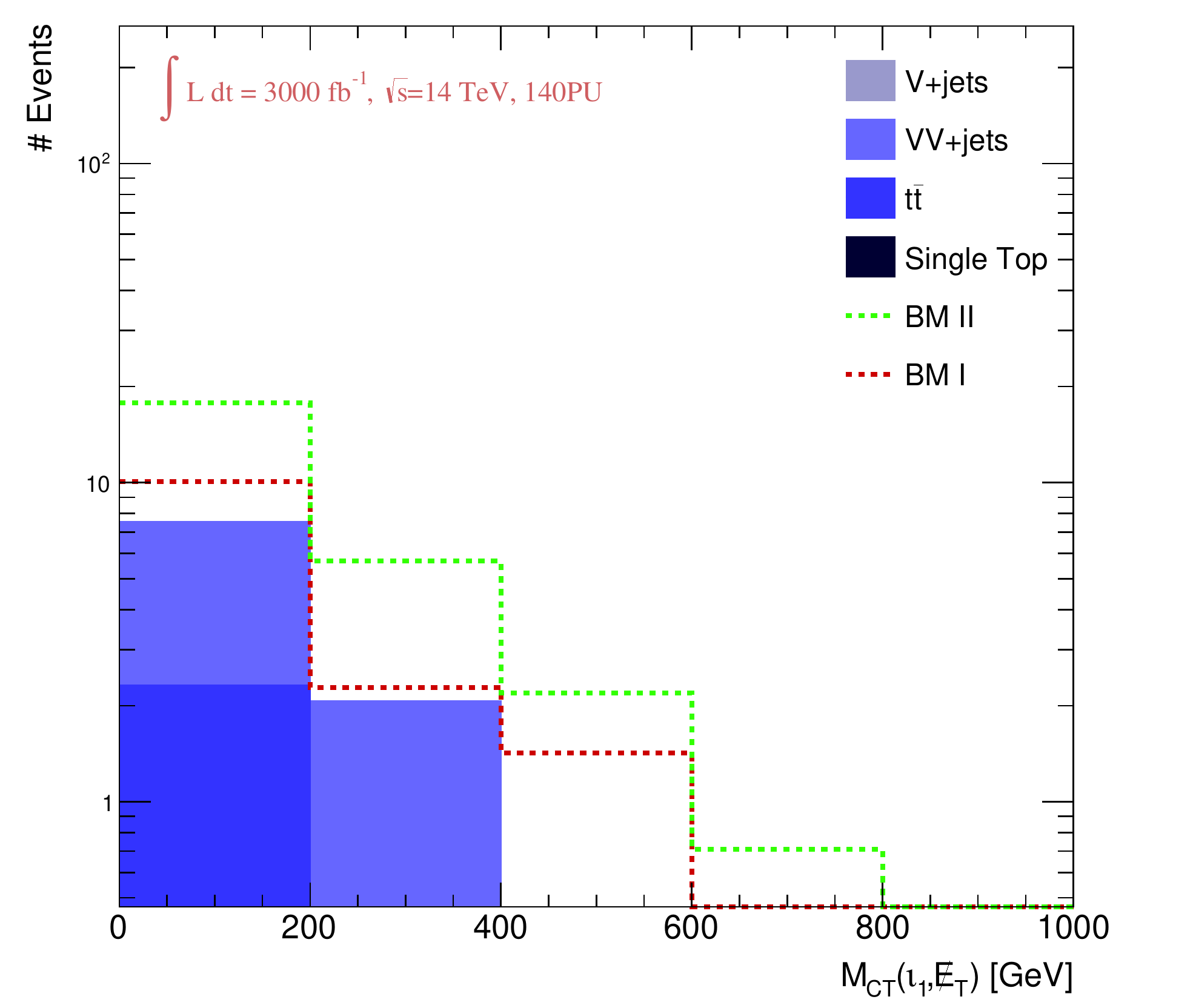}
 \caption{Left: missing transverse energy spectrum for the different components of
  the background and the two signal benchmarks.
  Right: cotransverse mass distributions for muon $l_1$ and
  invisible particles leading to missing energy
  (neutralinos and neutrinos).
  All histograms are obtained after applying all the acceptance cuts
  discussed in the paper.}
 \label{fig:mctmet}
\end{figure}

In the left panel of Fig.~\ref{fig:pt}, we present the distribution in the
transverse momentum of the leading muon $l_1$ after applying the first six cuts
of Table~\ref{tab:cutflow}. In the right panel of the figure, we in contrast
show the transverse-momentum spectrum of the next-to-leading muon $l_2$ as
resulting from the entire selection strategy.
As for the $p_T(l_1)$ spectrum, all four considered backgrounds
contribute at small $p_T$, while above 100 GeV the only surviving SM events
originate from the production of $VV$ and $t\bar t$ pairs.
The signal spectra are rather broad and lie below the
backgrounds at low $p_T(l_1)$, whereas, for
$p_T(l_1)>300$~GeV, both signals {\bf BM I} and {\bf BM II} start
to be competitive with the background, yielding comparable numbers of events.
For even larger transverse momenta, say $p_T(l_1)>$~500 GeV,
muons coming from supersymmetric decays of a leptophobic $Z^\prime$
become dominant, especially in the reference point {\bf BM II}.
After all cuts are applied, the $p_T(l_2)$
distribution is explored (Fig.~\ref{fig:pt}, right).
All backgrounds are further suppressed and those due to 
single vector-boson and single-top production are negligible.
The transverse momentum spectrum is thus substantial
in the $200~{\rm GeV}<p_T(l_2)<600~{\rm GeV}$ range,
with the {\bf BM II} signal yielding the highest number of events
through all $p_T$ range and {\bf BM I} being also quite remarkable,
especially for $200~{\rm GeV}<p_T<400~{\rm GeV}$.
Overall, Fig.~\ref{fig:pt} (right) shows that the
cuts which we have applied are rather efficient
to discriminate the leptons in leptophobic $Z^\prime$ events
from the Standard Model ones.

In Fig.~\ref{fig:mctmet} (left) we show the missing transverse energy,
due to the lightest neutralinos $\tilde\chi^0_1$ in our signal and
to neutrinos in the backgrounds, after all cuts are imposed.
The  $\slashed{E}_T$ spectra of our UMSSM benchmark scenarios are 
well above the backgrounds, once again limited to $VV$ and
$t\bar t$ pairs, through the whole $\slashed{E}_T$  range.
The {\bf BM II} configuration, in particular, is capable of
yielding a few events up to $\slashed{E}_T\simeq 600$~GeV,
while, above 400 GeV, the backgrounds are basically all suppressed.

We have verified that any other transverse observable, such as the $M_{T2}$ or
$M_{CT}$ variables defined in 
Refs.~\cite{Lester:1999tx,Barr:2003rg,Tovey:2008ui}, are not useful
for improving the considered selection strategy due to the too small mass
difference between the lightest chargino and the lightest neutralino. The main
features of the signal topology are in this case already captured by the
requirements on the lepton transverse momenta and on
the missing transverse energy.

This is illustrated in Fig.~\ref{fig:mctmet} (right),
where we present the cotransverse mass $M_{CT}$
distribution\footnote{Given two particles of transverse energies
 $E_{T,1}$ and $E_{T,2}$ and transverse momenta ${\vec p}_{T,1}$
 and ${\vec p}_{T,2}$, the cotransverse mass is defined as
 $M_{CT}^2 = (E_{T,1} + E_{T,2})^2 - ({\vec p}_{T,1}+{\vec p}_{T,2})^2$
 \cite{Tovey:2008ui}.} for the leading muon $l_1$ and all particles
contributing to the missing energy (lightest neutralinos
and neutrinos).
The $M_{CT}$ spectrum is 
qualitatively comparable to the 
$\slashed{E}_T$ one.
Both signals and backgrounds ($VV$ and $t\bar t$) peak at
similar values, although the number of events generated by
$Z^\prime$ decays is always larger than for SM processes, and for 
$M_{CT}>400$~GeV only signal events survive.
Designing an analysis with a possible extra cut on $M_{CT}$
would lead to a reduction in the
significance, as both $S$ and $B$ would be affected in the same way. Such a new
selection may, however, increase the sensitivity
for spectra featuring larger mass gaps. In this work, we nevertheless choose to
focus on the lighter UMSSM particle spectra that are still not excluded so far
and thus more relevant for the near future.

\section{Summary and Conclusions}
\label{sec:conclusion}
Motivated by the latest ATLAS and CMS measurements which
imposed improved lower bounds on the $Z^\prime$ mass, we analyzed models
with an additional  $U(1)^\prime$ gauge symmetry group arising from the
breaking of $E_6$
supersymmetric GUT. We explored
possible loopholes in the searches carried out
at the LHC. In particular, we allowed the $Z^\prime$ to decay into
supersymmetric final states, such as gaugino pairs, and investigated
scenarios where the $Z^\prime$ is leptophobic.
In fact, as the $Z^\prime$ mass bounds are mostly determined by its decay
into lepton pairs, the constraints would be relaxed in models in which
direct leptonic decays are suppressed or even forbidden.
We found that leptophobia can be achieved by accounting for the
kinetic mixing between the two $U(1)$ symmetries, parameterized by an angle
$\chi$, and that, 
among possible $U(1)^\prime$ groups, the model $U(1)^\prime_\eta$, while obeying all low energy conditions on the parameter space, is most favored to be leptophobic.
Our analysis was then undertaken under two possible assumptions for scale unification, the gauge couplings being imposed to unify either at the GUT scale or at $M_{Z^\prime}$.
We investigated the mass bounds and decay patterns in both cases,
as well as
the prospects for seeing a $Z^\prime$ signal above the background at the LHC,
accounting for supersymmetry and leptophobia.

Concerning supersymmetry, for both high- and 
low-scale unification, the rates of dilepton production are smaller once we include new decay modes, which translates into a reduction of the mass exclusion limits by about 200 GeV.
As for dijets, we found an even larger impact of the inclusion of
supersymmetric channels, so that the LHC constraints can be evaded.
Within leptophobic scenarios, observing supersymmetric $Z^\prime$
decays into charged leptons and missing energy would be most
promising through a cascade from a primary decay into chargino pairs.
We analyzed final-state
signals from these intermediate states and suppressed the background by imposing a jet veto,
in addition to requirements on the final-state leptons and missing energy.
We chose two benchmark points in the parameter space,
corresponding to different UMSSM realizations, and
found that they both yield visible signals at the LHC,
with a significance which varies from $3\sigma$ up to even
$7\sigma$, according to the criterion  employed to
estimate the LHC sensitivity. 

Therefore, supersymmetric and possibly leptophobic $Z^\prime$
decays are capable of giving detectable dilepton signals,
which can be easily discriminated from the backgrounds
and from non-supersymmetric $Z^\prime$ events, so far
employed to set the exclusion limits.
Moreover, from the viewpoint of supersymmetry, 
$Z^\prime$ bosons
would be a promising source of new particles, such as the charginos
and neutralinos investigated in this paper, which,
unlike direct production in $pp$ collisions,
would feature additional kinematic
constraints set by the high $Z^\prime$ mass. 

In summary, we believe that our study,
accounting
for Grand Unification Theories, supersymmetry and leptophobia
altogether, should 
represent a useful guiding reference to explore a more general
gauge structure than the Standard Model and 
address its 
incompleteness from perspectives that have
not received so far proper consideration from the experimental
collaborations. We  demonstrated that
investigating such scenarios is instead
both worthwhile and feasible, as they are 
potentially capable of giving remarkable signals, especially in the
high-luminosity phase of the LHC.

\begin{acknowledgments}
JYA thanks  Altan \c{C}ak\i r regarding the usage of the Snowmass background
samples, Florian Staub for the implementation of the model in SARAH and
\"{O}zg\"{u}r \c{S}ahin for providing help with the \textsc{ThePlotting}
software. MF acknowledges the NSERC for partial financial support under grant number SAP105354. The work of BF is partly supported by French state funds managed by the Agence Nationale de la Recherche (ANR), in the context of the LABEX ILP (ANR-11-IDEX-0004-02, ANR-10-LABX-63).
\end{acknowledgments}

\providecommand{\href}[2]{#2}\begingroup\raggedright\endgroup

\end{document}